\documentclass[12pt, a4paper]{article}

\usepackage[utf8]{inputenc}
\usepackage{amsmath, amsfonts, amssymb}
\usepackage{graphicx}
\usepackage{booktabs}
\usepackage{geometry}
\usepackage{hyperref}
\usepackage[numbers,square]{natbib}

\usepackage{verbatim}
\usepackage{setspace}
\usepackage{titling}
\usepackage{fancyhdr}
\usepackage{algorithm, algorithmic}
\usepackage{array}
\usepackage{multirow}
\usepackage{booktabs}

\usepackage{amssymb}   
\usepackage{amsmath}
\usepackage{amsthm}
\usepackage{mathtools}
\usepackage{enumitem}
\usepackage{subcaption}
\usepackage{pdfpages}
\usepackage{comment}

\title{\bfseries Taming the Greeks: \\ Option Portfolios with Inductive Biases}

\author{
    \small \textbf{Wee Ling Tan}\thanks{Corresponding author. Email: \texttt{weeling@robots.ox.ac.uk}. S. Roberts: \texttt{sjrob@robots.ox.ac.uk}. S. Zohren: \texttt{stefan.zohren@eng.ox.ac.uk}.} \\
    \small Department of Engineering Science \\ 
    \small Oxford-Man Institute of Quantitative Finance \\
    \small University of Oxford \\[2ex]
    \small \textbf{Stephen Roberts} \\
    \small Department of Engineering Science \\ 
    \small Oxford-Man Institute of Quantitative Finance \\
    \small University of Oxford \\[2ex]
    \small \textbf{Stefan Zohren} \\
    \small Department of Engineering Science \\ 
    \small University of Oxford
}
\date{}

\begin{document}

\maketitle
\vspace{-0.6cm}

\begin{abstract}
We present an end-to-end deep learning framework for systematic options trading that directly embeds hedging behavior through explicit control of portfolio-level risk exposures. While neural networks trained to optimize risk-adjusted performance have been shown to outperform traditional rules-based strategies, such approaches remain agnostic to the sensitivities of the resulting portfolios with respect to specific underlying risk factors. We propose a general training objective that combines a performance-driven loss with a differentiable risk-sensitivity penalty, enforcing neutrality to selected risk dimensions. Unlike reinforcement learning methods that approximate optimal hedging policies via simulated market dynamics, our framework operates entirely on historical data and jointly optimizes risk-adjusted returns and targeted risk constraints in a single learning problem. We instantiate the framework on static delta-neutral straddle portfolios with the penalty directed at first-order directional exposure, and evaluate two penalty variants -- an exposure-normalized penalty and a Greek-ratio drift penalty. Empirical results on Nasdaq 100 equity options demonstrate that appropriately calibrated regularization simultaneously improves out-of-sample risk-adjusted performance relative to an unregularized baseline while reducing realized directional exposure.
\end{abstract}


\section{Introduction}
\label{introduction}

Options are highly versatile financial instruments that provide investors the flexibility to express non-linear views on underlying assets. However, this flexibility comes at the cost of having to navigate a complex set of risk exposures. In addition to directional risks associated with the underlying asset, option portfolios are exposed to higher-order and exogenous risks such as convexity, implied volatility, time decay, and interest rates, typically summarized by the Greeks \cite{hull2016options}. Executing systematic options strategies often requires strict adherence to these risk constraints. In particular, institutional trading desks must dynamically manage portfolio risk exposures in order to isolate targeted risk premia.

Managing these exposures has traditionally relied on model-based hedging techniques grounded in assumptions about the underlying market dynamics and option pricing models. Recent literature has increasingly turned to data-driven methodologies, including the use of machine learning algorithms to construct options trading strategies \cite{tan2024deep}. While these models often exhibit strong empirical performance, optimization objectives that focus solely on maximizing risk-adjusted returns can result in strategies that implicitly load on specific undesirable risks. We argue that risk metrics directly observable from the options chain provide an informative representation of the portfolio's state at any given time, and should therefore be integrated into the signal generation process. As such, our approach explicitly incorporates risk-sensitivity measures into the learning objective, effectively constraining the portfolio's exposure to specific risk dimensions.

Concretely, in this work we propose a novel optimization framework for systematic options trading strategies that embeds penalties for portfolio-level risk sensitivities directly into the learning process. The resulting training procedure drives the generated trading signals towards neutrality, or selectively control, with respect to specified risk dimensions, while jointly optimizing risk-adjusted returns. Our approach differs fundamentally from reinforcement learning methods for hedging, which rely on simulating different market paths to approximate dynamic hedging policies  \cite{buehler2019deep, kolm2019dynamic}. In contrast, our framework operates entirely on empirical historical data, enabling the joint optimization of performance and targeted risk constraints without requiring a generative model of market dynamics.

A natural starting point is to couple a performance-driven objective with a regularization term that penalizes a selected Greek exposure of the portfolio. However, because portfolio-level Greek exposures scale linearly with position sizes, such a penalty can lead to a degenerate solution in which the model learns the trivial solution by uniformly scaling down trading positions. Consequently, the strategy converges to a form of regularization collapse in which the portfolio appears well hedged only because trading activity has been effectively suppressed. To address this limitation, we introduce a reformulation of the regularization objective to penalize inefficient risk composition rather than absolute exposure levels. Firstly, we introduce a position-normalized penalty that computes Greek exposure per unit of gross signal, decoupling the risk term from the overall scale of positions. Secondly, we introduce a drift penalty defined through a Greek-to-Greek ratio that identifies contracts whose risk profile has become dominated by a particular risk sensitivity. We develop and evaluate these objectives on static delta-neutral straddle portfolios, a structure whose risk profile is concentrated around the at-the-money (ATM) region and thus provides a controlled setting for studying the interaction between regularization and directional exposure.

Our framework preserves the main advantages of end-to-end optimization while introducing an explicit mechanism for controlling risk exposures during training. To illustrate this, we systematically train neural networks subject to varying degrees of regularization and analyze the resulting risk profiles of the options portfolio over time. We demonstrate that appropriately calibrated models not only exhibit competitive out-of-sample hedging performance but also simultaneously improve risk-adjusted returns relative to an unregularized baseline. We further extend our framework to incorporate turnover regularization, optimizing for net-of-fee performance metrics in the presence of transaction costs.

\section{Related Work}

The theoretical foundations of option pricing is fundamentally linked to the seminal work of Black-Scholes-Merton \cite{black1973pricing, merton1973theory}, establishing the principle of no-arbitrage pricing under the assumption that an option's payoff can be perfectly replicated using a continuously rebalanced portfolio of the underlying asset and a risk-free bond. Subsequent work extended these ideas to more flexible and data-driven settings. For example, \cite{hutchinson1994nonparametric} demonstrate that neural networks can approximate option pricing functions directly from market data, while more recent studies use non-parametric approaches to model option prices \cite{ivașcu2021option}.

Since the introduction of option pricing models, a significant volume of literature has focused on the problem of optimal replication and hedging. The works of \cite{boyle1992option, figlewski1989options, leland1985option} address the concept of discrete time hedging in the presence of transaction costs. More recently, reinforcement learning methods have been proposed for dynamic hedging under realistic trading constraints. \cite{buehler2019deep, kolm2019dynamic} formulate hedging as a sequential decision-making problem and learn hedging policies for derivative portfolios using simulated market paths, followed by several other works extending this framework \cite{gao2023deeper, hirano2023adversarial, mueller2024fast}. These approaches are primarily motivated by the objectives of market makers and liquidity providers, who seek to minimize hedging costs and manage risk across large books of derivatives. In contrast, our work considers the problem from the perspective of an active investor constructing systematic trading strategies for options portfolios. Rather than learning policies through simulating market dynamics, we directly optimize trading signals using historical market data. In addition, we explicitly incorporate portfolio-level risk sensitivities into the learning objective, jointly maximizing risk-adjusted performance while controlling risk exposures.

Although much of the existing literature centers on the pricing and hedging of derivatives, another line of research examines systematic risk premia and pricing anomalies in options markets \cite{buchner2022factor, coval2001expected}. These findings are coupled with the development of systematic trading strategies aimed at exploiting such premia \cite{goyal2009cross, vasquez2017equity}, including momentum and trend-following strategies \cite{heston2023option}. Broadly, trend-based strategies fall into two complementary paradigms: momentum strategies, which exploit the tendency of asset returns to persist in the same direction \cite{jegadeesh1993returns, moskowitz2012time}, and mean-reversion strategies, which take contrarian positions anticipating the correction of overextended trends \cite{de1985does, poterba1988mean}. Our work contributes primarily to this body of literature, focusing on systematic trend-following strategies within options markets. A key advantage of these strategies is their simplicity and reliance on directly observable option prices and returns, circumventing the need to impose specific option pricing models a priori. Some studies adopt a decoupled, two-stage framework: for instance, \cite{bali2023option} use both linear and nonlinear models with option- and stock-specific inputs to predict returns of delta-hedged options, subsequently forming portfolios using the forecasted returns. In contrast, our approach proposes a single, end-to-end function that concurrently handles trend prediction and optimal position sizing. In \cite{tan2024deep}, the authors demonstrate that end-to-end deep learning frameworks substantially outperform conventional rules-based strategies when trading the cross-section of S\&P 100 equity options over an out-of-sample period spanning more than a decade. Rather than attempting to predict asset returns, neural networks are trained to directly map normalized point-in-time option features into target position weights. This unified architecture addresses a fundamental limitation in predict-then-optimize approaches as shown by \cite{lim2019enhancing}, where highly accurate return forecasts do not necessarily translate into profitable strategies. As \cite{harvey2018impact} note, the overall profitability of a systematic trading strategy is heavily influenced by dynamic factors that are strictly exogenous to the predictive model, such as the position sizing of portfolio positions, incorporation of risk constraints, and the impact of transaction costs.

While end-to-end models can efficiently maximize aggregate portfolio metrics such as risk-adjusted returns \cite{lim2019enhancing, tan2023spatio, wood2021trading, zhang2020deep}, institutional trading desks are often strictly bound by risk limits and thresholds, therefore necessitating the integration of risk-constraint mechanisms. Our current framework addresses this challenge by directly embedding risk-constraint objectives into the learning objective, enabling varying degrees of risk control to be incorporated during training.

\section{Dataset Description}
We obtain our equity option dataset from the OptionMetrics Ivy DB database. For each contract, the database provides end-of-day bid and ask quotes, implied volatilities (IVs), and option sensitivities (Greeks). The IVs and Greeks are calculated with a binomial tree model using Cox, Ross, and Rubinstein \cite{cox1979option}. We restrict our analysis to the universe of option contracts written on constituents of the Nasdaq 100 Index over the period from January 2010 to December 2023. This universe comprises a broad cross-section of large-cap US equities and are characterized by relatively deep and liquid options markets.

Given the large number of contracts in our universe, we apply a series of data filters consistent with \cite{goyal2009cross, heston2023option, tan2024deep}. First, we retain only standard monthly options expiring on the third Friday of each month, excluding weeklies and contracts affected by special settlements or other corporate actions. We further eliminate observations that violate the arbitrage bounds of American-style options, observations that exhibit zero bid prices, or whose ask price is less than or equal to the bid. Finally, only contracts with strictly positive open interest at the portfolio formation date are retained in order to exclude options with no liquidity.

On each monthly expiration date (hereafter referred to as the portfolio formation date), we construct a static delta-neutral straddle for each stock by pairing a call and a put with the same strike price, both expiring in the following month. Our sample targets options that are closest to at-the-money, i.e. we identify the call-put pair whose moneyness (calculated as $S/K$ for calls and $K/S$ for puts, where $S$ is the stock price and $K$ is the strike price) lies closest to 1.0, restricting the selection to contracts within a moneyness interval of 0.95 to 1.05. Equivalently, this selection criterion targets a straddle Delta of close to zero at initiation.

\section{Problem Definition}
\subsection{Portfolio Formation and Returns Computation}
In this section, we formalize the problem of managing a portfolio of options within a systematic strategy. Consider a portfolio of straddle options, where $i = 1, 2, \cdots, N_t$ indexes the individual underlying stocks. We compute the overall daily strategy returns as follows:

\begin{align}
    R_{t+1}^{\Pi} &= \frac{1}{N_t} \sum_{i=1}^{N_t} X_{i,t} \left( \frac{\sigma^*}{\sigma_{i,t}} \right) R_{i, t+1} \label{eqn:strategy_return}
\end{align}
where individual straddle returns are defined as:
\begin{align}
    R_{i, t+1} &= \frac{V_{i, t+1} - V_{i, t}}{V_{i, t}} \label{eqn:straddle_return} \\
    V_{i, t} &= w_{i}^C C_{i,t} + w_{i}^P P_{i,t} \label{eqn:straddle_value} 
\end{align}
with respective weights:
\begin{equation}
    w_{i}^C = \frac{-\Delta_{i,0}^P}{\Delta_{i,0}^C - \Delta_{i,0}^P}, \qquad w_{i}^P = \frac{\Delta_{i,0}^C}{\Delta_{i,0}^C - \Delta_{i,0}^P} \label{eqn:straddle_weights}
\end{equation}

Given each stock $i$, we initiate a delta-neutral straddle position on the day of portfolio formation ($t=0$). The position is established by assigning pre-normalized weights of $- \Delta_{i,0}^P$ and $\Delta_{i,0}^C$ to the respective call and put options, where $ \Delta_0$ denotes the option's initial delta. After normalization, these weight allocations remain fixed until expiration. This static hedging approach is consistent with \cite{goyal2009cross, heston2023option, tan2024deep}, where a one-time delta hedge is implemented only at initiation, with no subsequent adjustments to the weights. To compute the daily returns of the straddle, we track the combined value of the constituent options using their closing bid-ask midpoints. This approach to computing returns inherently assumes European-style execution, given our focus on near-ATM options that retain substantial extrinsic value. Following \cite{moskowitz2012time}, we apply volatility scaling on the trading signal $X_{i,t} \in [-1, 1]$ of the individual straddle options as defined in Equation (\ref{eqn:strategy_return}) by targeting a constant annualized volatility, $\sigma^*$. The ex-ante volatility, $\sigma_{i,t}$, is estimated using a 20-day exponentially weighted moving standard deviation of the straddle returns.

\begin{subsection}{Systematic Strategies}
\label{systematic_strategies}
All systematic options portfolio strategies require, at least, an estimate of $X_{i,t}$, which specifies the trading position for a particular straddle option at any given time. Since our interest is in studying the risk-adjusted performance of systematic trend-based strategies, we construct several of these portfolios and calculate their returns based on Equation (\ref{eqn:strategy_return}):

\begin{enumerate}[label=(\arabic*)]
    \item \textbf{Long Only.} This strategy maintains a maximum long position for all straddle options in the portfolio, such that $X_{i,t}= 1$ for all $i$ and $t$.
    \item \textbf{Time Series Momentum (TSMOM).} The seminal work of \cite{moskowitz2012time} documents evidence of time-series momentum across a broad range of asset classes. In its simplest form, the standard specification for the momentum trading signal is binary ($-1$ or $+1$), based on the sign of the $n$-month asset returns. Consistent with \cite{heston2023option, tan2024deep}, we adopt the strategy with a monthly lookback period where $n = 1$. The trading signal is therefore given by $X_{i,t} = \text{sgn}(R_{t-20, t})$ where $R_{t-20, t}$ denotes the rolling 20-day return of the straddle option.
    \item \textbf{Moving Average Convergence Divergence (MACD).} We consider a more complex trend estimation function for the trading signal $X_{i,t}$ based on \cite{baz2015dissecting} that uses volatility-normalized, multi-horizon MACD indicators in place of the sign of returns for estimating the trading signal. At its core, the MACD captures price trends by measuring the distance between a fast and slow moving average. We compute MACD signals as an equally weighted sum over multiple short and long time scales $S \in \{2,4,8\}$ and $L \in \{8,16,32\}$. Full implementation details are provided in \cite{tan2024deep}.
    \item \textbf{Heston Option Momentum (TSHestonMOM, CSHestonMOM).} Unlike classical time-series momentum strategies which exploit serial dependence in the returns of a continuously traded asset, the strategy formalized by \cite{heston2023option} relies on cross-serial predictability across successive generations of option contracts, utilizing returns of past, expired straddle options to formulate trading signals for the currently traded portfolio of contracts. To implement this, we analyze the historical performance of 1-month, delta-neutral ATM straddles. At portfolio formation ($t=0$) and for a given stock $i$, we calculate its rolling $n$-month average straddle returns, denoted as $\bar{R}_{i, -n, 0}$, across four distinct lookback periods: $n \in \{1, 3, 6, 12\}$ months, generating the raw continuous signal $Y_{i,0} = \bar{R}_{i, -n, 0}$. For the TSHestonMOM strategy, we map the sign of this raw continuous signal to the trading signal, $X_{i,t} = \text{sgn}(Y_{i,0})$, which dictates the static position held to expiry. On the other hand, we construct a long-short decile portfolio on the day of portfolio formation for the CSHestonMOM strategy, ranking stocks by their raw option momentum signals $Y_{i,0}$. We refer the reader to \cite{heston2023option} for further details on the specific implementation.

\end{enumerate}

For all benchmarks, we also evaluate the inverse allocation, $-X_{i,t}$, which translates to a contrarian, mean-reverting (MR) strategy.

\subsection{Deep Learning Models}

Rather than constructing trading rules from a priori specified rules, we consider how machine learning methods can be used to model the trading signal $X_{i,t}$. The central task is therefore to determine how positions in the individual straddle options should be allocated as a function of the information available on each trading day.

Each straddle option, written on stock $i$ at time $t$, is characterized by a $d$-dimensional vector of observable features, $\mathbf{u}_{i, t}$. This feature vector is provided as the input to a parameterized decision function $f(\cdot)$, which models the trading signal $X_{i,t}$. Formally, we have:

\begin{equation}
\label{eqn:general_function}
X_{i, t} = f_{\boldsymbol{\theta}}
\left(
\mathbf{u}_{i, t}
\right)
\end{equation}

Under this specification, the model observes the present state as characterized by the input feature vector and is trained to learn a profitable end-to-end trading policy. This framework enables specific objectives and inductive biases to be embedded during training according to the portfolio manager's requirements, ultimately contributing to the final trading rule. We detail several of these training objectives in Section \ref{learning_to_hedge}. Consequently, the tasks of trend prediction and position sizing, which are typically treated separately, are now unified within a single optimization problem. Following \cite{tan2024deep}, we take the end-to-end model $f(\cdot)$ to be the Long Short-term Memory (LSTM) architecture and refer the reader to the text for further details on the specific implementation of the deep learning model.

\end{subsection}

\begin{subsubsection}{Predictor Description}
Following Equation (\ref{eqn:general_function}), we formulate the feature tensor $\mathbf{u}_{i, t}$ by incorporating a combination of the predictors associated with the systematic trend-following strategies detailed in Section \ref{systematic_strategies}, alongside a set of risk-sensitivity features required by the optimization step. This specific feature set fulfills two main objectives: it ensures direct comparability of our methodology with the trend-following benchmarks, and it provides a controlled framework to evaluate the marginal benefits of incorporating risk-sensitivity metrics during training.

The specific predictors are:

\begin{enumerate}[label=(\arabic*)]
    \item \textbf{Normalized Returns} -- we take historical straddle returns that are normalized over multiple time horizons. Specifically, the return over a given interval is scaled by the daily volatility estimate multiplied by $\sqrt{k}$, expressed as $R_{i, t-k, t} \slash (\sigma_{i, t} \sqrt{k})$. By computing this risk-adjusted metric for rolling windows of $k \in \{1, 5, 10, 15, 20\}$ trading days, we capture the multi-horizon dynamics of the strategy.
    
    \item \textbf{MACD Indicators} -- we incorporate the volatility-adjusted MACD signals, $Y_{i, t} (S, L)$, as defined in \cite{tan2024deep}. These are generated by applying a combination of fast and slow moving average windows, where $S \in \{2, 4, 8\}$ and $L \in \{8, 16, 32\}$, respectively.
  
    \item \textbf{Heston Option Momentum} -- as detailed in Section \ref{systematic_strategies} and \cite{heston2023option}, we incorporate option momentum characteristics into our feature set by averaging the target stock's straddle returns across trailing intervals of $n \in \{1, 3, 6, 12\}$ months.
    
    \item \textbf{Other Option Features} -- we include only two time-varying metrics: the log-moneyness of the individual put and call options, and the annualized days to expiry (DTE). These inputs characterize the option's specific phase within its lifecycle following the portfolio formation date, and are directly observable features.

    \item \textbf{Risk-Sensitivity Measures (Greeks)} -- we include risk-sensitivity measures, specifically the Greeks, of the static delta-neutral straddle. These are computed according to the weight allocation at initiation and serve to capture the evolving non-linear risk exposures that naturally fluctuate over the life of the trade. Specifically, we incorporate the standard set of option Greeks provided by the OptionMetrics dataset: Delta, Gamma, Theta, Vega, Rho. While our study utilizes risk-sensitivity measures sourced directly from this dataset, incorporating these predictors does not strictly tie our framework to any singular option pricing model. Our framework is designed to be inherently model-agnostic, allowing for the integration of alternative risk metrics evaluated under an arbitrary option pricing model.
    
\end{enumerate}

\end{subsubsection}

\section{Learning to Hedge with Inductive Bias}
\label{learning_to_hedge}
\subsection{General Framework}

We propose a hybrid training objective $\mathcal{L}(\boldsymbol{\theta})$ that combines a performance-driven objective with an explicit risk-sensitivity penalty. The resulting framework enables the construction of option trading strategies that embed hedging behavior at the loss level through the optimization process:

\begin{equation}
\label{general_objective_function}
\mathcal{L}(\boldsymbol{\theta}) = \mathcal{J}(\boldsymbol{\theta}) + \alpha \| \nabla_{\mathbf{x}} \mathcal{V} \|
\end{equation}
where $\mathcal{J}(\boldsymbol{\theta})$ denotes a general performance-driven objective and $\| \nabla_{\mathbf{x}} \mathcal{V} \|$ defines a risk-sensitivity measure of the portfolio value ($\mathcal{V}$) with respect to specific underlying factors ($\mathbf{x}$), whose specific functional form is defined in Section \ref{subsec:risk_sensitivity_penalty}. Here, the scalar $\alpha \geq 0$ serves as a risk-aversion coefficient that governs the trade-off between the base performance metric and the hedging penalty. In this study, we designate $\alpha$ as a tunable hyperparameter and calibrate the model's tolerance for risk exposure, gradually shifting the optimization trade-off from maximizing the performance-driven metric ($\alpha \to 0$) towards enforcing strict penalties on risk sensitivity.

\subsection{Performance-Driven Objective}
Let $\mathcal{D}_{\Omega} = \left\{ \left( \mathbf{u}_{i, t}, X_{i, t} \right) \right\}$ represent the empirical dataset pairing observed market states with the corresponding trading signals generated by the network, where $X_{i, t} = f_{\boldsymbol{\theta}}(\mathbf{u}_{i, t})$. Here, the set $\Omega$ indexes all $N_t$ available straddles across $T$ time steps. While this framework supports any differentiable performance metric, for the remainder of this paper, we define $\mathcal{J}(\boldsymbol{\theta})$ to maximize risk-adjusted returns by setting it to the negative expected Sharpe ratio \cite{sharpe1998sharpe} evaluated over $\mathcal{D}_{\Omega}$: 

$$\mathcal{J}(\boldsymbol{\theta}) = - \sqrt{252} \left( \frac{\mathbb{E}_{\Omega}[\tilde{R}_{i, t+1}]}{\sqrt{\text{Var}_{\Omega}(\tilde{R}_{i, t+1})}} \right),$$
$$\tilde{R}_{i, t+1} = X_{i, t} \left(\frac{\sigma^*}{\sigma_{i, t}}\right) R_{i, t+1}.$$

\subsection{Risk-Sensitivity Penalty}
\label{subsec:risk_sensitivity_penalty}
Having established the generalized objective function in Equation (\ref{general_objective_function}), we now specify the functional form of the risk-sensitivity norm, $\| \nabla_{\mathbf{x}} \mathcal{V} \|$. The choice of functional form is motivated by the fundamental risk dynamics of straddle portfolios, where for a small perturbation in the underlying risk factor, the local variation in portfolio value can be approximated by a discrete Taylor expansion with respect to the primary option risk factors \cite{estrella1997approximation}:

\begin{equation}
d\mathcal{V}
\approx
\Delta \, dS
+
\frac{1}{2}\Gamma (dS)^2
+
\nu \, d\sigma
+
\Theta \, dt.
\end{equation}
where
\begin{equation}
\Delta = \frac{\partial \mathcal{V}}{\partial S},
\qquad
\Gamma = \frac{\partial^2 \mathcal{V}}{\partial S^2},
\qquad
\nu = \frac{\partial \mathcal{V}}{\partial \sigma},
\qquad
\Theta = \frac{\partial \mathcal{V}}{\partial t}.
\end{equation}

Here, Delta ($\Delta$) captures first-order directional exposure, Gamma ($\Gamma$) captures curvature with respect to the underlying asset price, Vega ($\nu$) captures sensitivity to implied volatility, and Theta ($\Theta$) captures sensitivity to the passage of time. For example, these sensitivities are evaluated analytically under the Black–Scholes–Merton framework \cite{black1973pricing, merton1973theory}, which provides closed-form expressions for the Greeks of European-style options. This decomposition provides a natural interpretation for the general risk-sensitivity norm $\| \nabla_{\mathbf{x}} \mathcal{V} \|$: by selecting the components of $\mathbf{x}$, the penalty can be directed towards specific sources of portfolio risk. While the general risk-sensitivity term is, in principle, capable of regularizing an arbitrary set of option risk exposures, including Gamma, Vega, Theta, or higher-order sensitivities, the empirical scope of our work specifically targets the first-order directional exposure, namely Delta. 

At inception, the at-the-money straddle is explicitly constructed to be directionally neutral ($\Delta \approx 0$). At this stage, the variance of the strategy's payoff is driven primarily by second-order convexity (Gamma) and volatility exposure (Vega). However, as the underlying asset price drifts away from the initial strike, the risk properties of the options fundamentally change. Gamma decays towards zero as convexity decreases, while the net Delta accumulates towards an absolute value of 1.0 as one leg moves deep in-the-money. Consequently, the first-order linear term ($\Delta dS$) comes to dominate the second-order term, and unhedged Delta dominates the variance of the strategy's payoff. Without explicit regularization, the straddle ceases to be directionally neutral and acquires a substantial amount of directional sensitivity.

Our focus on Delta in this work provides a controlled setting in which to study whether the risk-sensitivity penalty can embed hedging behavior directly into the training objective. By isolating a single, interpretable source of risk, we remove the confounding effects that may arise when multiple Greeks are penalized simultaneously. Extensions to multi-dimensional risk-sensitivity penalties are therefore left for future work. Consequently, for the remainder of this paper, we restrict the market factor vector $\mathbf{x}$ to a single variable: the underlying asset price, $S$. Under these conditions, the gradient norm mathematically simplifies to the absolute magnitude of the portfolio's Delta, allowing us to formalize the specific penalty variants in the following sections.

By construction, the risk-sensitivity norm $\| \nabla_{\mathbf{x}} \mathcal{V} \|$ is defined solely in terms of partial derivatives of the portfolio value function and is therefore independent of the option pricing model used to evaluate these derivatives. The Greeks employed in our experiments are adopted as a practical default given their availability in the dataset. Substituting risk measures derived from an alternative pricing model therefore requires no modification to the objective function or optimization procedure.

\subsubsection{Na\"{i}ve Risk-Sensitivity Penalty}

A direct implementation of the risk-sensitivity term would penalize the absolute Delta exposure of the portfolio. This corresponds to an L1-type penalty of the form:
\begin{equation}
\| \nabla_{\mathbf{x}} \mathcal{V} \|_{\text{naive}} = \sum_{(i,t)\in\Omega}
\left| X_{i,t}\Delta_{i,t} \right|,
\end{equation}
where $X_{i,t}$ denotes the allocation to straddle $i$ at time $t$, and $\Delta_{i,t}$ denotes its first-order sensitivity with respect to the underlying asset price. Although this specification directly penalizes directional exposure, it admits a degenerate solution. Since the penalty is also directly proportional to the absolute allocation size, the optimizer can reduce the risk term simply by shrinking all trading signals towards zero. We refer to this failure mode as signal shrinkage.

To maintain active trading signals while discouraging undesirable directional exposure, we propose and evaluate two distinct penalty variants. The first normalizes directional exposure by the magnitude of the total allocation, producing an exposure-normalized risk measure. The second uses a Greek-ratio weighting scheme that penalizes allocations to contracts whose Greeks indicate substantial drift away from the at-the-money region.

\subsubsection{Exposure-Normalized Penalty}

The first variant normalizes the portfolio's gross directional exposure by the total magnitude of the model's allocations. This converts the penalty into a per-unit-allocation measure of directional risk, thereby limiting the optimizer's tendency to reduce the penalty solely by scaling down all positions. Given the index set $\Omega$ of all straddle-time observations, we define the exposure-normalized Delta penalty as:

\begin{equation}
\| \nabla_{\mathbf{x}} \mathcal{V} \|_{\text{ENP}} =
\frac{
\sum_{(i,t)\in\Omega}
\left| X_{i,t}\Delta_{i,t} \right|
}{
\sum_{(i,t)\in\Omega}
\left| X_{i,t} \right|
+
\epsilon
}
\end{equation}
where $\epsilon>0$ is a small regularizing constant introduced for numerical stability. The numerator measures the gross directional exposure induced by the model's allocations, while the denominator rescales this exposure by the total amount of trading activity. As a result, the penalty evaluates whether the strategy is directionally exposed relative to its own allocation, removing the incentive to reduce the risk term by uniformly scaling down positions.

We also consider a quadratic L2 extension of this normalized penalty:

\begin{equation}
\| \nabla_{\mathbf{x}} \mathcal{V} \|_{\text{ENP-quad}} = \frac{
\sum_{(i,t)\in\Omega}
( X_{i,t}\Delta_{i,t} )^2
}{
\sum_{(i,t)\in\Omega}
( X_{i,t} )^2
+
\epsilon
}
\end{equation}

This amplifies the penalty assigned to large aggregate directional exposures while retaining normalization by the total squared allocations. By introducing convexity into the exposure term, the quadratic variant places disproportionate weight on concentrated unhedged Delta risk, rather than penalizing all trading activity uniformly. Empirically, this further mitigates signal shrinkage by discouraging excessive directional exposure without creating a blanket incentive for the model to suppress all trading signals.

\subsubsection{Drift Penalty (Greek Ratio)}
We propose a second variant that uses the relationship between Greeks to identify contracts that have drifted away from the region where the straddle convexity is most pronounced. We define the drift penalty as:

\begin{equation}
\| \nabla_{\mathbf{x}} \mathcal{V} \|_{\text{DP}} =
\frac{
\sum_{(i,t)\in\Omega}
\left| X_{i,t}\Delta_{i,t} \right|
}{
\sum_{(i,t)\in\Omega}
\left| X_{i,t}\Gamma_{i,t} \right|
+
\epsilon
}
\end{equation}
Similar to the exposure-normalized approach, we extend the drift penalty by applying a quadratic transformation to the realized directional exposure. By squaring the numerator while normalizing against the squared realized convexity in the denominator, we define the quadratic drift penalty as:

\begin{equation}
\| \nabla_{\mathbf{x}} \mathcal{V} \|_{\text{DP-quad}} = \frac{
\sum_{(i,t)\in\Omega}
(X_{i,t} \Delta_{i,t})^2
}{
\sum_{(i,t)\in\Omega}
(X_{i,t} \Gamma_{i,t})^2
+
\epsilon
}
\end{equation}

For straddle portfolios, Gamma is typically largest near the at-the-money region, while Delta exposure accumulates as the underlying asset price moves away from the initial strike. The ratio ($|\Delta_{i,t}/\Gamma_{i,t}|$) therefore provides a local measure of whether a position is dominated by directional exposure relative to its curvature. The drift penalty assigns a larger cost to allocations in contracts whose Delta is large relative to Gamma where such positions are more likely to behave as directional exposures. The loss function therefore discourages the model from maintaining positions in straddles that have drifted too far from the at-the-money region, while still permitting exposure to contracts whose curvature remains large relative to their directional sensitivity. Unlike the exposure-normalized penalty, which evaluates realized Delta per unit of allocation, the drift penalty acts as a structural risk weight on the opportunity set itself. It therefore discourages the model from allocating capital to contracts whose local Greek profile indicates that the straddle has become directionally dominated. This provides a complementary mechanism for stabilizing unhedged straddle portfolios while preserving active signal generation.

\subsection{Optimization}
We minimize the training objectives via minibatch stochastic gradient descent with the Adam optimizer \cite{kingma2014adam}. For each in-sample window, the available data is split chronologically into a training set (first 90\%) and a validation set (final 10\%). We use the latter exclusively for early stopping, halting training if no improvement in the validation loss is observed for 25 consecutive epochs. Hyperparameter selection is conducted through a random search over 100 candidate configurations, with full details of the search ranges in Appendix~\ref{appendix_a}. All experiments were carried out on a server with an AMD EPYC7713 CPU and multiple NVIDIA L40 GPUs.

\subsection{Backtest Details}
To evaluate out-of-sample performance, we employ an expanding window approach with five-year increments: each successive training window extends by an additional five years of data, after which the calibrated model, with all parameters and hyperparameters held fixed, is evaluated on the subsequent five-year period. We repeat this procedure across several independently seeded trials to account for variability arising from random initialization. The out-of-sample performance from all seeds are aggregated and the combined results are presented in Section~\ref{results_and_discussion}.

\section{Results and Discussion}
\label{results_and_discussion}

In this section, we structure our analysis into three primary objectives. First, we evaluate the out-of-sample performance of all systematic strategies. Second, we examine whether incorporating risk-sensitivity penalties into the learning objective reduces the realized directional exposure of the learned portfolios. Third, we analyze the trade-off induced by varying the risk-aversion coefficient ($\alpha$), in particular, whether stronger regularization improves hedging stability at the cost of risk-adjusted performance. Lastly, we analyze the impact of transaction costs on the risk-adjusted performance of each strategy.

\subsection{Out-of-Sample Strategy Performance}
\label{subsec:oos_performance}

\begin{table*}[htbp]
\centering
\caption{Performance Metrics - Rescaled to Target Volatility}
\label{table:best_models}
\resizebox{\textwidth}{!}{
\begin{tabular}{lccccccccc}
\hline \toprule
\textbf{Model} & \textbf{E[Return]} & \textbf{Vol.} & \textbf{\begin{tabular}[c]{@{}l@{}} Downside \\ Deviation \end{tabular}} & \textbf{MDD} & \textbf{Sharpe} & \textbf{Sortino} & \textbf{Calmar} & \textbf{\begin{tabular}[c]{@{}l@{}} Hit \\ Rate \end{tabular}} & \textbf{ $\mathbf{\frac{\text{Ave. P}}{\text{Ave. L}}}$ } \\
\midrule
\multicolumn{10}{l}{\underline{\textbf{Benchmarks}}} \\
Long Only & 0.100 & 0.161 & 0.088 & 0.340 & 0.621 & 1.129 & 0.294 & 0.447 & 1.390 \\
TSMR & 0.132 & 0.161 & \textbf{0.086} & 0.311 & 0.822 & 1.537 & 0.426 & 0.449 & 1.431 \\
MACDMR & 0.128 & 0.163 & \textbf{0.086} & 0.276 & 0.785 & 1.487 & 0.465 & 0.447 & \textbf{1.440} \\
TSHestonMR & 0.117 & \textbf{0.156} & 0.100 & 0.268 & 0.747 & 1.166 & 0.435 & 0.494 & 1.168 \\
CSHestonMR & 0.118 & 0.158 & 0.104 & 0.292 & 0.742 & 1.135 & 0.403 & 0.511 & 1.098 \\
\midrule
\multicolumn{10}{l}{\underline{\textbf{Baseline}}} \\
Baseline & 0.399 & 0.187 & 0.131 & 0.222 & 2.131 & 3.043 & 1.797 & 0.656 & 1.053 \\
\midrule
\multicolumn{10}{l}{\underline{\textbf{Regularized Models}}} \\
ENP (L1) & 0.422 & 0.187 & 0.130 & 0.225 & 2.255 & 3.259 & 1.878 & 0.662 & 1.090 \\
ENP (L2) & 0.409 & 0.181 & 0.129 & 0.236 & 2.260 & 3.167 & 1.735 & 0.638 & 1.182 \\
DP (L1) & \textbf{0.436} & 0.186 & 0.122 & \textbf{0.210} & \textbf{2.339} & \textbf{3.583} & \textbf{2.078} & 0.675 & 1.101 \\
DP (L2) & 0.425 & 0.188 & 0.131 & 0.246 & 2.255 & 3.236 & 1.724 & \textbf{0.681} & 1.012 \\
\bottomrule \hline
\end{tabular}
}
\begin{flushleft}$_{\text{\ \ \ \ (bold denotes best performing strategy for each column)}}$\end{flushleft}
\vfill
\end{table*}

We begin by comparing the out-of-sample risk-adjusted performance of all strategies computed using the overall returns as defined in Equation (\ref{eqn:strategy_return}). We evaluate the performance of three groups of systematic strategies: traditional rules-based benchmarks (Benchmarks), a deep learning model trained solely on the Sharpe objective (Baseline), and deep learning models trained with the proposed objective that includes a risk-sensitivity penalty (Regularized Models). At the portfolio level, we rescale all signals to target an annualized volatility of $15\%$ and report the out-of-sample performance of each strategy in Table \ref{table:best_models}. Since momentum-based strategies were generally unprofitable over the out-of-sample period, we report only the inverse allocation (MR) strategies for the benchmarks. For the Heston portfolios (TSHestonMR, CSHestonMR), we report results for the best performing lookback period for brevity. In this section, we report results in the absence of transaction costs to assess the raw predictive ability of each strategy. The impact of transaction costs is addressed in Section \ref{subsec:transaction_cost_impact}, where we conduct an analysis on signal turnover.

We use the following annualized performance metrics:

\begin{enumerate}[label = \textbf{\arabic*.}]
\item \textbf{Profitability Metrics} -- Expected Return (($\mathbb{E}[\text{Returns}]$)) and Hit Rate.

\item \textbf{Risk Metrics} -- Volatility (Vol.), Downside Deviation, and Maximum Drawdown (MDD).

\item \textbf{Performance Ratios} -- Sharpe Ratio, Sortino Ratio, Calmar Ratio, and Average Profit-over-Loss $\left(\frac{\text{Ave. P}}{\text{Ave. L}}\right)$

\end{enumerate}

From Table~\ref{table:best_models}, we observe that the benchmark strategies: Long Only, TSMR, MACDMR, TSHestonMR, and CSHestonMR deliver annualized Sharpe ratios in the range of 0.621 to 0.822. While these strategies generate positive returns, their risk-adjusted performance remains modest. The baseline deep learning model, which is trained to maximize only the performance-driven objective as in Equation (\ref{general_objective_function}), in this case the Sharpe ratio without a risk-sensitivity penalty ($\alpha = 0$), achieves a substantially higher out-of-sample Sharpe ratio of 2.131. This confirms the effectiveness of end-to-end optimization for learning profitable trading signals, consistent with \cite{tan2024deep}, and establishes the relevant performance benchmark against which we evaluate the regularized models. Although the baseline model achieves substantially stronger risk-adjusted performance than the benchmarks, we show in Section \ref{subsec:delta_exposure} this performance is accompanied by a persistent positive net Delta bias and comparatively high gross Delta exposure.

We conduct controlled experiments varying the strength of the penalties of the regularized models, and report only the best performing model for each regularized variant in this section for brevity. We refer the reader to Section \ref{subsec:risk_return_tradeoff} for a full analysis of these controlled experiments. Notably, we see that the regularized models not only preserve but improve upon the baseline model's risk-adjusted performance. Among the regularized variants, the exposure-normalized penalty (ENP) models achieve a Sharpe ratio of 2.260, while the drift penalty (DP) models attain the highest Sharpe ratio of 2.339 with the lowest maximum drawdown across all models. We observe that the incorporation of risk-sensitivity regularization does not inherently penalize performance. Instead, at appropriately calibrated levels, several regularized variants were observed to attain stronger risk-adjusted performances compared to the baseline.

\subsection{Realized Portfolio Delta Exposure}
\label{subsec:delta_exposure}

To assess the directional exposure of each strategy, we evaluate two complementary, position-normalized measures of the portfolio's end-of-day Delta. For a given day $t$, the first evaluation measure is the net position-normalized Delta:

\begin{equation}
\label{eqn:evaluation_net_position_normalized_delta}
\bar{\Delta}^{\mathrm{net}}_t
=
\frac{\sum_{i=1}^{N_t} X_{i,t}\Delta_{i,t}}
{\sum_{i=1}^{N_t}|X_{i,t}|}
\end{equation}
where $\Delta_{i,t}$ is the corresponding end-of-day straddle Delta. Equation~\eqref{eqn:evaluation_net_position_normalized_delta} represents the portfolio's net directional exposure per unit of gross allocation. Since positive and negative contributions enter with their respective signs, this measure reflects the aggregate Delta tilt of the combined portfolio, but this can be reduced through cross-sectional offsetting.

To distinguish reductions in the magnitude of the underlying exposures from such offsetting, we also report the gross position-normalized Delta:

\begin{equation}
\label{eqn:evaluation_gross_position_normalized_delta}
\bar{\Delta}^{\mathrm{gross}}_t
=
\frac{\sum_{i=1}^{N_t}|X_{i,t}\Delta_{i,t}|}
{\sum_{i=1}^{N_t}|X_{i,t}|}
\end{equation}
where this measure captures the position-normalized average absolute Delta of all positions held at the end-of-day by the model. By definition, since individual exposures enter in absolute value, no cancellation between positive and negative Delta contributions occurs, and the metric therefore reflects the gross directional exposure embedded in the individual positions of the strategy.

Together, these two measures allow us to jointly analyze the extent to which individual strategies achieve directional neutrality: whether through cross-sectional netting at the portfolio level or through hedging at the level of individual positions. For example, a portfolio may exhibit near-zero net Delta while retaining substantial gross Delta if its constituent positions carry large but offsetting directional exposures. The joint analysis of net and gross Delta exposures therefore provides a richer characterization of the risk exposure of the strategy than either measure alone.

Importantly, both measures are invariant to a uniform rescaling of all positions on a given day. Neither metric can be reduced solely by shrinking all trading signals by a common factor -- changes in any of these normalized metrics would require a different cross-sectional allocation across the portfolio or, for the net measure, a different balance between positive and negative contributions.

We note that both of these metrics aggregate Delta contributions across positions written on different underlying stocks. Since each straddle option is constructed to be delta-neutral at inception, the individual Deltas $\Delta_{i,t}$ are on a comparable scale -- each begins near zero and drifts as the underlying moves away from the initial strike. These metrics therefore provide a meaningful cross-sectional summary of how much directional exposure the portfolio has accumulated per unit of gross allocation, and should be interpreted as measures of average position-level directional drift rather than the  portfolio's sensitivity to any single underlying.

\subsubsection{Net Position-Normalized Delta}
\label{subsubsec:net_delta}

\begin{figure*}[htbp]
    \centering
    \includegraphics[width=\textwidth]{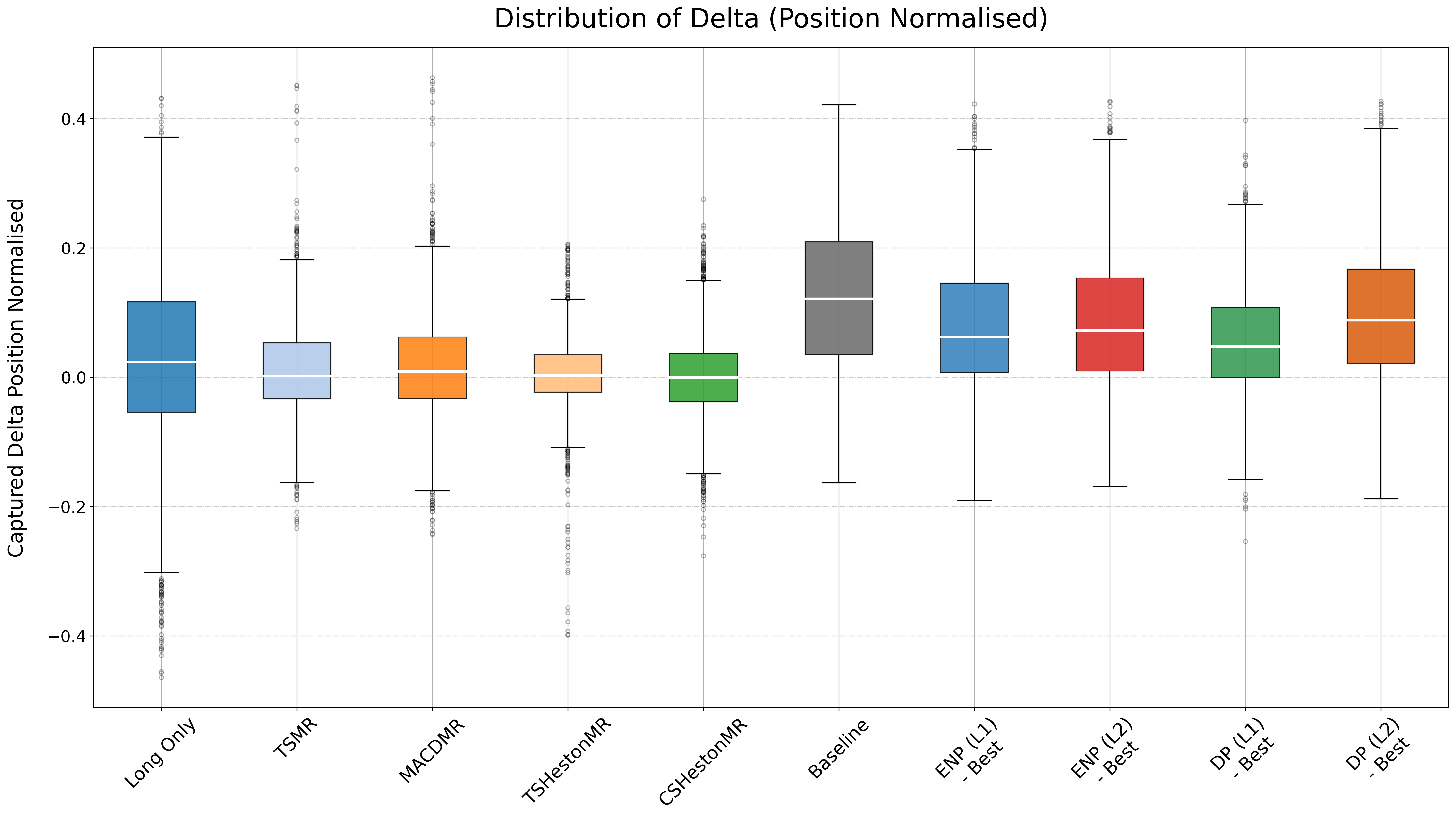}
    \caption{Distribution of Daily Net Position-Normalized Delta across the Out-of-Sample Period}
    \label{fig:position_normalized_delta_distribution}
\end{figure*}

\begin{table*}[htbp]
\centering
\caption{Summary Statistics: Net Position-Normalized Delta}
\label{tab:summary_stats_selected_models_position_normalised_delta}
\resizebox{\textwidth}{!}{
\begin{tabular}{lccccccc}
\hline \toprule
\textbf{Model} & \textbf{Mean} & \textbf{Std} & \textbf{Min} & \textbf{Q1} & \textbf{Median} & \textbf{Q3} & \textbf{Max} \\
\midrule
\multicolumn{8}{l}{\underline{\textbf{Benchmarks}}} \\
Long Only       & 0.023 & 0.142 & -0.465 & -0.054 &  0.024 & 0.117 & 0.432 \\
TSMR            & 0.011 & 0.076 & -0.234 & -0.033 &  0.002 & 0.053 & 0.452 \\
MACDMR          & 0.015 & 0.084 & -0.243 & -0.033 &  0.009 & 0.062 & 0.464 \\
TSHestonMR      & 0.004 & 0.061 & -0.398 & -0.023 &  0.003 & 0.035 & 0.206 \\
CSHestonMR      & 0.001 & 0.067 & -0.276 & -0.038 & -0.000 & 0.037 & 0.276 \\
\midrule
\multicolumn{8}{l}{\underline{\textbf{Baseline}}} \\
Baseline        & 0.127 & 0.107 & -0.164 &  0.035 &  0.121 & 0.210 & 0.421 \\
\midrule
\multicolumn{8}{l}{\underline{\textbf{Regularized Models}}} \\
ENP (L1) & 0.083 & 0.099 & -0.190 &  0.007 &  0.062 & 0.146 & 0.423 \\
ENP (L2) & 0.089 & 0.098 & -0.169 &  0.010 &  0.072 & 0.153 & 0.427 \\
DP (L1)  & 0.056 & 0.081 & -0.254 & -0.000 &  0.047 & 0.108 & 0.397 \\
DP (L2)  & 0.101 & 0.099 & -0.188 &  0.021 &  0.088 & 0.167 & 0.427 \\
\bottomrule \hline
\end{tabular}
}
\end{table*}

We plot the distribution of daily aggregate portfolio net position-normalized Delta over the entire out-of-sample period in Figure~\ref{fig:position_normalized_delta_distribution} and provide the summary statistics in Table~\ref{tab:summary_stats_selected_models_position_normalised_delta}. 

First, we observe that the baseline model learns a pronounced positive directional exposure, with a mean daily net position-normalized Delta of $0.127$ and median at $0.121$. Since even the first quartile is positive, the central mass of the baseline model's net Delta distribution is shifted above zero rather than being driven by a small number of positive outliers. The baseline model therefore appears to be exposed to persistent positive Delta as part of its return-generating policy, despite holding a portfolio of delta-neutral straddles at formation. This confirms that, in the absence of explicit risk constraints, the end-to-end optimization process implicitly loads on Delta exposure when maximizing the performance-driven objective.

The benchmark strategies are more closely centered around zero in net terms. Long Only, TSMR, MACDMR, TSHestonMR, and CSHestonMR have mean net Deltas between $0.001$ and $0.023$. Their relatively neutral aggregate exposures, however, do not necessarily imply that the constituent straddle positions are uniformly better hedged, since positive and negative Delta contributions can offset. Next, we observe that the distribution of net position-normalized Delta of the Long Only model has a mean Delta of $0.023$ but a large standard deviation of $0.142$, while TSMR, MACDMR, TSHestonMR, and CSHestonMR have medians close to zero and narrower distributions. However, the Sharpe ratios of all benchmark strategies are below $0.9$. The central question is therefore not whether the net Deltas of each strategy can be mechanically centered around zero, but whether incorporating risk-sensitivity penalties in the training objective of the baseline model can lead to a reduction in the net directional bias of the learned portfolio while retaining the performance gains of end-to-end learning.

We observe that across all regularized variants, both the ENP and DP models produce a reduction in realized net Delta exposure as compared to the baseline, with mean net Deltas between $0.056$ and $0.101$ as compared to $0.127$ for the baseline model. In addition, we see a similar reduction in median net Deltas. These results show that each of these variants can lead to a reduction in net aggregate directional exposure. In the next section, we examine whether this reduction reflects lower absolute Delta exposures at the position level.

\subsubsection{Gross Position-Normalized Delta}
\label{subsubsec:gross_delta}

\begin{figure*}[htbp]
    \centering
    \includegraphics[width=\textwidth]{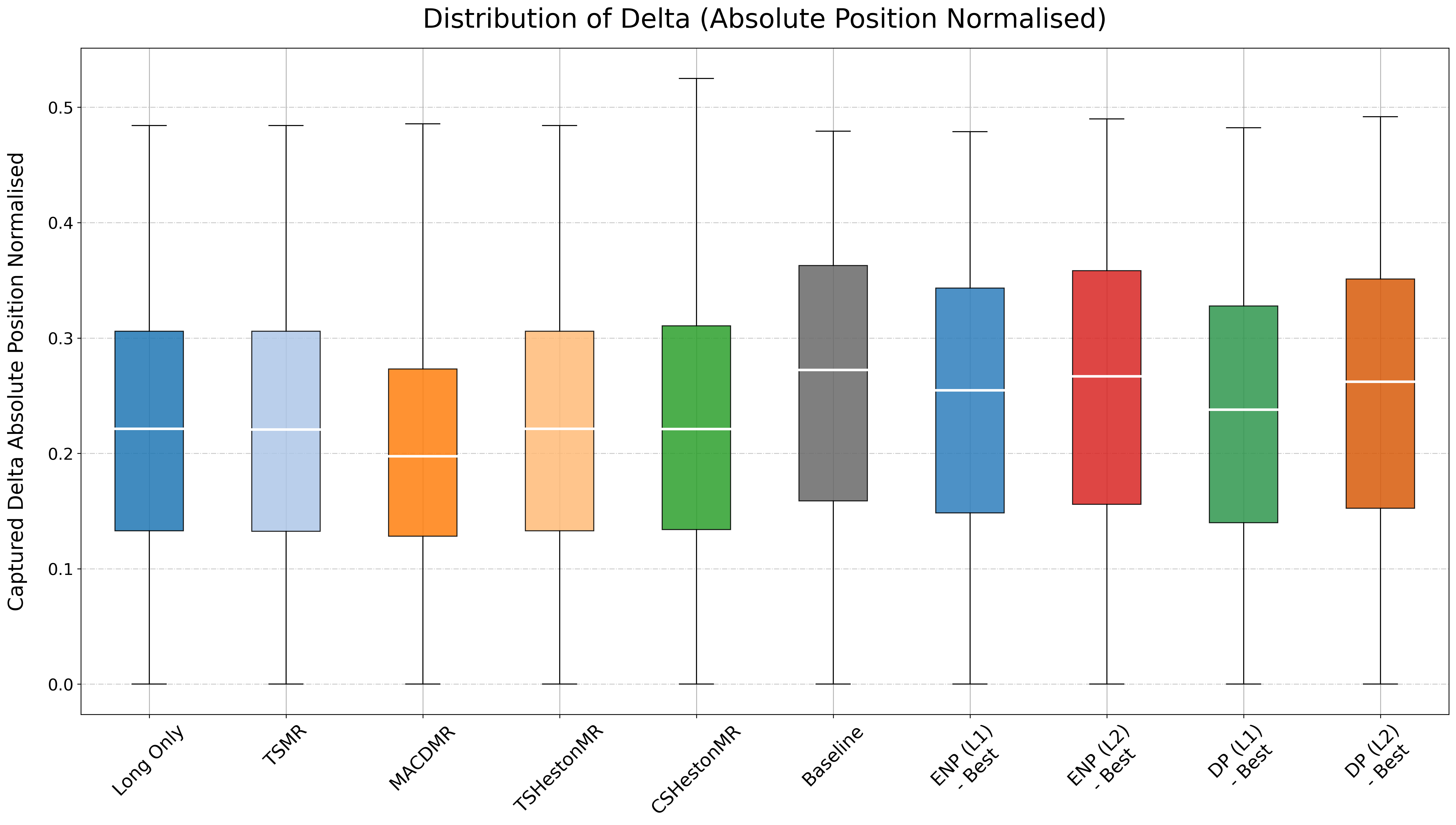}
    \caption{Distribution of Daily Gross Position-Normalized Delta across the Out-of-Sample Period}
    \label{fig:absolute_position_normalized_delta_distribution}
\end{figure*}

\begin{table*}[htbp]
\centering
\caption{Summary Statistics: Gross Position-Normalized Delta}
\label{tab:summary_stats_selected_models_absolute_position_normalised_delta}
\resizebox{\textwidth}{!}{
\begin{tabular}{lccccccc}
\hline \toprule
\textbf{Model} & \textbf{Mean} & \textbf{Std} & \textbf{Min} & \textbf{Q1} & \textbf{Median} & \textbf{Q3} & \textbf{Max} \\
\midrule
\multicolumn{8}{l}{\underline{\textbf{Benchmarks}}} \\
Long Only       & 0.218 & 0.110 & 0.000 & 0.133 & 0.221 & 0.306 & 0.484 \\
TSMR            & 0.218 & 0.110 & 0.000 & 0.132 & 0.221 & 0.306 & 0.484 \\
MACDMR          & 0.201 & 0.101 & 0.000 & 0.128 & 0.198 & 0.273 & 0.486 \\
TSHestonMR      & 0.218 & 0.110 & 0.000 & 0.133 & 0.221 & 0.306 & 0.484 \\
CSHestonMR      & 0.219 & 0.113 & 0.000 & 0.134 & 0.221 & 0.311 & 0.525 \\
\midrule
\multicolumn{8}{l}{\underline{\textbf{Baseline}}} \\
Baseline        & 0.256 & 0.123 & 0.000 & 0.159 & 0.272 & 0.363 & 0.479 \\
\midrule
\multicolumn{8}{l}{\underline{\textbf{Regularized Models}}} \\
ENP (L1) & 0.243 & 0.120 & 0.000 & 0.148 & 0.255 & 0.343 & 0.479 \\
ENP (L2) & 0.252 & 0.123 & 0.000 & 0.156 & 0.267 & 0.358 & 0.490 \\
DP (L1)  & 0.231 & 0.116 & 0.000 & 0.140 & 0.238 & 0.328 & 0.482 \\
DP (L2)  & 0.247 & 0.121 & 0.000 & 0.152 & 0.262 & 0.351 & 0.492 \\
\bottomrule \hline
\end{tabular}
}
\end{table*}

Figure~\ref{fig:absolute_position_normalized_delta_distribution} and Table~\ref{tab:summary_stats_selected_models_absolute_position_normalised_delta} provide a stricter assessment of hedging effectiveness at the position level. The baseline model has a mean gross position-normalized Delta of $0.256$ and a median of $0.272$. These values are materially larger than its net mean and median, confirming that the portfolio contains substantial Delta exposures whose positive and negative components partially offset in aggregation. The benchmark strategies have gross means between $0.201$ and $0.219$, lower than the baseline model but also far from zero. Thus, their near-zero net Deltas can be primarily attributed to portfolio-level netting rather than an absence of realized Delta at the position level.

The DP (L1) model, the best-performing variant overall, saw a slight reduction in the gross mean Delta exposure from $0.256$ to $0.231$ and a reduction in the median from $0.272$ to $0.238$, despite a larger decline in net Delta exposure. Similarly, the ENP models realize a slightly lower mean and median gross Delta compared to the baseline. These are meaningful but considerably smaller improvements than those implied by the net Delta measure, and these values are largely similar to the benchmark strategies. 

Therefore, the risk-sensitivity penalties seem to act primarily by reducing the baseline model's persistent positive bias and improving the balance and netting of Delta contributions at the portfolio level, while only moderately reducing the gross Delta exposure at the position level.

\subsection{The Risk--Return Trade-off Across Penalty Strengths}
\label{subsec:risk_return_tradeoff}

\begin{figure*}[t]
    \centering
    \includegraphics[width=\textwidth]{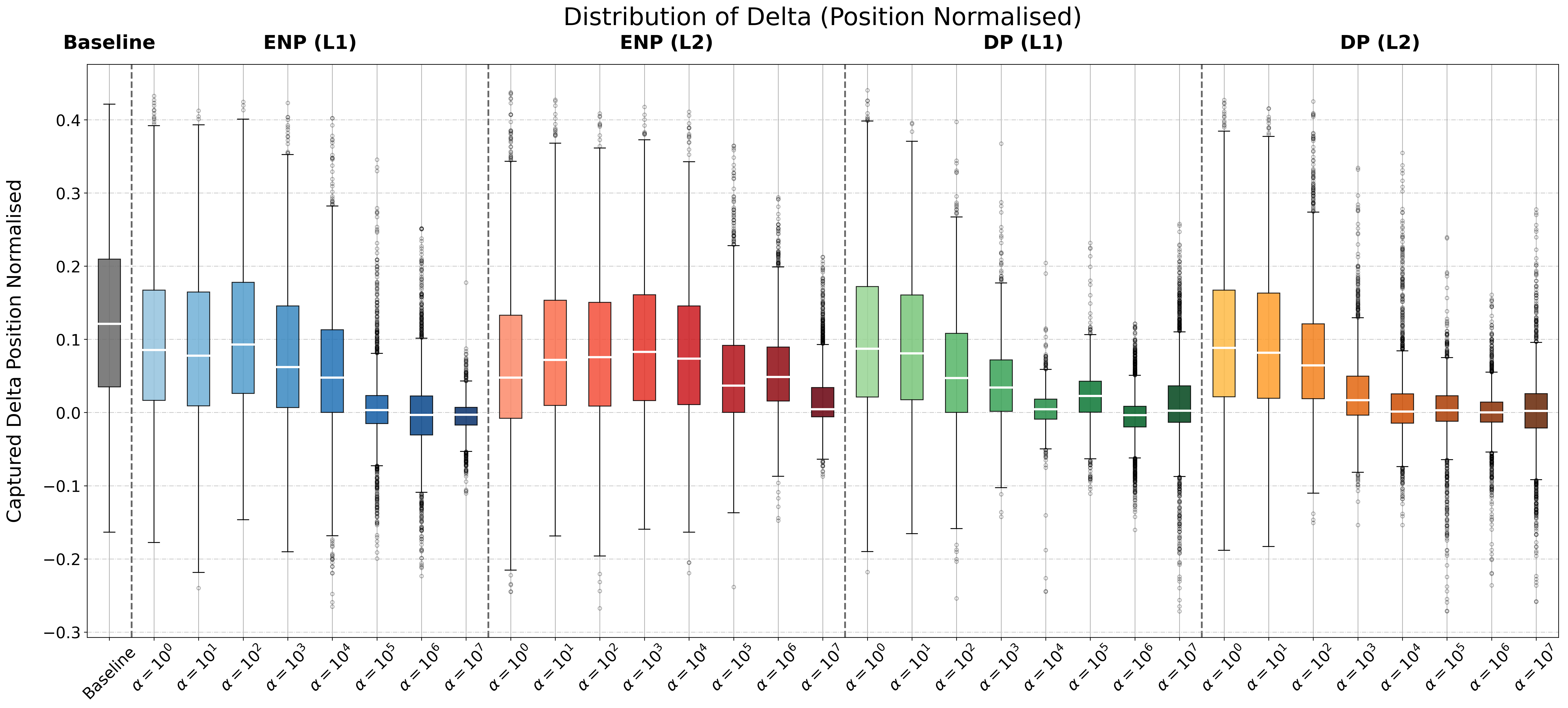}
    \caption{Distribution of Daily Net Position-Normalized Delta across all Penalty Strengths}
    \label{fig:net_position_normalized_delta_distribution_dlmodelsonly}
\end{figure*}

\begin{figure*}[t]
    \centering
    \includegraphics[width=\textwidth]{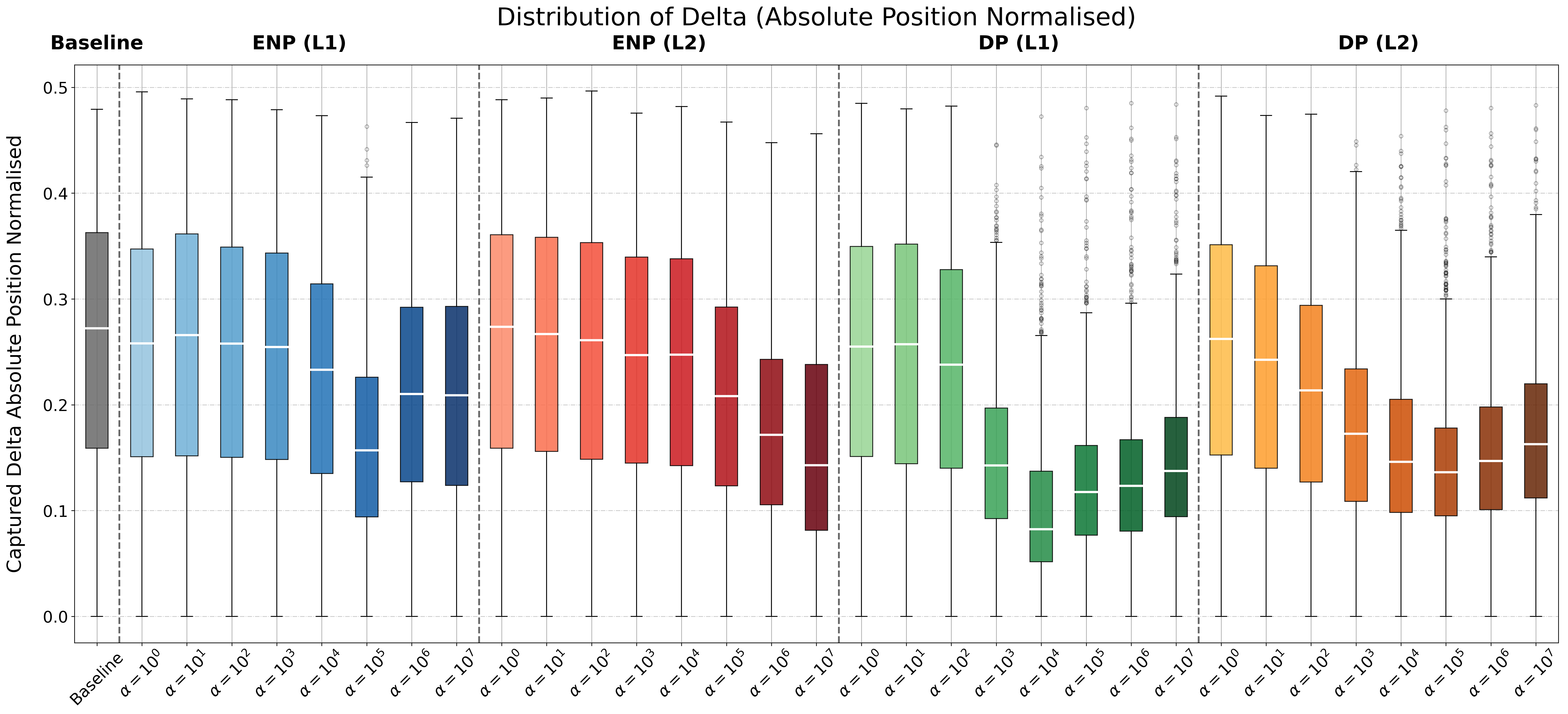}
    \caption{Distribution of Daily Gross Position-Normalized Delta across all Penalty Strengths}
    \label{fig:gross_position_normalized_delta_distribution_dlmodelsonly}
\end{figure*}

\begin{table*}[htbp]
\centering
\caption{Performance Metrics: Baseline vs. Regularized Models - Rescaled to Target Volatility}
\label{table:regularized_vs_baseline}
\resizebox{\textwidth}{!}{
\begin{tabular}{lccccccccc}
\hline \toprule
\textbf{Model} & \textbf{E[Return]} & \textbf{Vol.} & \textbf{\begin{tabular}[c]{@{}l@{}} Downside \\ Deviation \end{tabular}} & \textbf{MDD} & \textbf{Sharpe} & \textbf{Sortino} & \textbf{Calmar} & \textbf{\begin{tabular}[c]{@{}l@{}} Hit \\ Rate \end{tabular}} & \textbf{ $\mathbf{\frac{\text{Ave. P}}{\text{Ave. L}}}$ } \\
\midrule
\multicolumn{10}{l}{\underline{\textbf{Baseline}}} \\
Baseline & 0.399 & 0.187 & 0.131 & 0.222 & 2.131 & 3.043 & 1.797 & 0.656 & 1.053 \\
\midrule
\multicolumn{10}{l}{\underline{\textbf{Exposure-Normalized Penalty (L1)}}} \\
ENP (L1) $\alpha=10^{0}$ & 0.365 & 0.199 & 0.144 & 0.283 & 1.833 & 2.527 & 1.290 & 0.655 & 1.020 \\
ENP (L1) $\alpha=10^{1}$ & 0.415 & 0.196 & 0.141 & 0.268 & 2.113 & 2.941 & 1.549 & 0.671 & 1.034 \\
ENP (L1) $\alpha=10^{2}$ & 0.417 & 0.187 & 0.128 & 0.255 & 2.231 & 3.260 & 1.632 & 0.661 & 1.125 \\
ENP (L1) $\alpha=10^{3}$ & 0.422 & 0.187 & 0.130 & 0.225 & 2.255 & 3.259 & 1.878 & 0.662 & 1.090 \\
ENP (L1) $\alpha=10^{4}$ & 0.367 & 0.197 & 0.139 & 0.245 & 1.865 & 2.638 & 1.496 & 0.674 & 0.965 \\
ENP (L1) $\alpha=10^{5}$ & 0.247 & 0.231 & 0.157 & 0.420 & 1.070 & 1.574 & 0.587 & 0.600 & 1.450 \\
ENP (L1) $\alpha=10^{6}$ & 0.122 & 0.182 & 0.146 & 0.427 & 0.667 & 0.831 & 0.285 & 0.557 & 1.570 \\
ENP (L1) $\alpha=10^{7}$ & 0.097 & 0.171 & 0.141 & 0.405 & 0.567 & 0.688 & 0.239 & 0.478 & \textbf{1.868} \\
\midrule
\multicolumn{10}{l}{\underline{\textbf{Exposure-Normalized Penalty (L2)}}} \\
ENP (L2) $\alpha=10^{0}$ & 0.394 & 0.195 & 0.139 & 0.277 & 2.024 & 2.838 & 1.426 & 0.668 & 1.016 \\
ENP (L2) $\alpha=10^{1}$ & 0.409 & 0.181 & 0.129 & 0.236 & 2.260 & 3.167 & 1.735 & 0.638 & 1.182 \\
ENP (L2) $\alpha=10^{2}$ & 0.400 & 0.190 & 0.133 & 0.274 & 2.107 & 3.016 & 1.459 & 0.677 & 1.008 \\
ENP (L2) $\alpha=10^{3}$ & 0.400 & 0.192 & 0.131 & 0.238 & 2.088 & 3.057 & 1.679 & 0.643 & 1.178 \\
ENP (L2) $\alpha=10^{4}$ & 0.380 & 0.189 & 0.127 & 0.236 & 2.007 & 2.984 & 1.612 & 0.679 & 0.937 \\
ENP (L2) $\alpha=10^{5}$ & 0.349 & 0.194 & 0.123 & 0.246 & 1.794 & 2.825 & 1.419 & 0.636 & 1.205 \\
ENP (L2) $\alpha=10^{6}$ & 0.288 & 0.244 & 0.173 & 0.390 & 1.179 & 1.668 & 0.737 & 0.614 & 1.308 \\
ENP (L2) $\alpha=10^{7}$ & 0.072 & 0.181 & 0.130 & 0.441 & 0.397 & 0.553 & 0.162 & 0.553 & 1.033 \\
\midrule
\multicolumn{10}{l}{\underline{\textbf{Drift Penalty (L1)}}} \\
DP (L1) $\alpha=10^{0}$ & 0.397 & 0.185 & 0.128 & 0.219 & 2.139 & 3.098 & 1.814 & 0.657 & 1.056 \\
DP (L1) $\alpha=10^{1}$ & 0.414 & 0.191 & 0.133 & 0.235 & 2.161 & 3.120 & 1.758 & 0.679 & 1.008 \\
DP (L1) $\alpha=10^{2}$ & \textbf{0.436} & 0.186 & 0.122 & \textbf{0.210} & \textbf{2.339} & \textbf{3.583} & \textbf{2.078} & 0.675 & 1.101 \\
DP (L1) $\alpha=10^{3}$ & 0.362 & 0.207 & 0.145 & 0.251 & 1.753 & 2.506 & 1.442 & 0.647 & 1.123 \\
DP (L1) $\alpha=10^{4}$ & 0.145 & 0.155 & 0.122 & 0.316 & 0.935 & 1.193 & 0.460 & 0.529 & 1.425 \\
DP (L1) $\alpha=10^{5}$ & 0.042 & 0.104 & 0.089 & 0.285 & 0.402 & 0.470 & 0.147 & 0.471 & 1.377 \\
DP (L1) $\alpha=10^{6}$ & -0.012 & 0.112 & 0.098 & 0.433 & -0.106 & -0.121 & -0.027 & 0.562 & 0.727 \\
DP (L1) $\alpha=10^{7}$ & 0.020 & 0.151 & 0.125 & 0.460 & 0.129 & 0.156 & 0.042 & 0.580 & 0.779 \\
\midrule
\multicolumn{10}{l}{\underline{\textbf{Drift Penalty (L2)}}} \\
DP (L2) $\alpha=10^{0}$ & 0.425 & 0.188 & 0.131 & 0.246 & 2.255 & 3.236 & 1.724 & \textbf{0.681} & 1.012 \\
DP (L2) $\alpha=10^{1}$ & 0.382 & 0.195 & 0.134 & 0.215 & 1.954 & 2.840 & 1.778 & 0.669 & 1.050 \\
DP (L2) $\alpha=10^{2}$ & 0.386 & 0.195 & 0.128 & 0.237 & 1.977 & 3.012 & 1.630 & 0.660 & 1.142 \\
DP (L2) $\alpha=10^{3}$ & 0.353 & 0.189 & 0.111 & 0.212 & 1.865 & 3.189 & 1.669 & 0.651 & 1.239 \\
DP (L2) $\alpha=10^{4}$ & 0.124 & 0.164 & 0.097 & 0.298 & 0.759 & 1.285 & 0.417 & 0.524 & 1.257 \\
DP (L2) $\alpha=10^{5}$ & 0.054 & 0.132 & 0.071 & 0.238 & 0.414 & 0.766 & 0.229 & 0.464 & 1.386 \\
DP (L2) $\alpha=10^{6}$ & -0.019 & \textbf{0.066} & \textbf{0.048} & 0.213 & -0.294 & -0.405 & -0.091 & 0.480 & 0.966 \\
DP (L2) $\alpha=10^{7}$ & 0.001 & 0.146 & 0.112 & 0.412 & 0.009 & 0.012 & 0.003 & 0.499 & 1.010 \\
\bottomrule \hline
\end{tabular}
}
\end{table*}

In this section, we conduct a sensitivity analysis to study the effect of varying the power of the risk-sensitivity penalties of the regularized model, analyzing the trade-offs induced by systematically varying the risk-aversion coefficient ($\alpha$). This coefficient explicitly controls the trade-off between the performance-driven objective and the risk-sensitivity penalty.

First, referring to Table \ref{table:regularized_vs_baseline}, the relationship between $\alpha$ and the out-of-sample performance of the strategy is non-monotonic. At moderate levels, the penalties frequently improve risk-adjusted performance, indicating that the risk-sensitivity term can serve as a beneficial inductive bias rather than purely as a Delta constraint. For example, the models ENP (L2) at $\alpha=10^1$, DP (L1) at $\alpha=10^2$, and DP (L2) at $\alpha=10^0$ either improve or preserve downside-risk metrics while outperforming the baseline based on the Sharpe ratio.

Second, referring to Figures \ref{fig:net_position_normalized_delta_distribution_dlmodelsonly} and \ref{fig:gross_position_normalized_delta_distribution_dlmodelsonly}, we observe the distinction between the resulting net and gross Delta exposures of the strategy. The regularized models generally reduce net Delta more sharply than gross absolute Delta. This difference suggests that part of the hedging improvement comes from eliminating the baseline model's persistent positive Delta tilt at the portfolio level and balancing positive and negative Delta contributions across the portfolio. Such cross-sectional netting reduces the aggregate first-order directional sensitivity at the portfolio level. However, it is not equivalent to selecting individual positions with uniformly low absolute Delta. The gross Delta metric therefore provides a more demanding measure of whether the model is exposed to high Delta straddles at the position level.

Third, the choice of penalty norm influences the shape of the trade-off. The DP variants can produce sharper reductions in net and gross Delta, but their risk-adjusted performance rapidly deteriorates once the regularization coefficient crosses a variant-specific threshold. This is especially visible for DP (L1), where the Sharpe ratio falls from $2.339$ at $\alpha=10^2$ to $0.935$ at $\alpha=10^4$, and for DP (L2), whose Sharpe ratio falls from $2.255$ at $\alpha=10^0$ to $0.759$ at $\alpha=10^4$. In contrast, we observe that the ENP variants produce more gradual exposure reductions and retain stronger performance over a wider range of the regularization coefficient $\alpha$. ENP (L2), in particular, retains a Sharpe ratio above $2.0$ across $\alpha=10^0$ through $10^4$, while eventually reducing the median net Delta to $0.072$.

Fourth, focusing our analysis on net Delta, no single model specification dominates on every objective. DP (L1) at a level of $\alpha=10^2$ outperforms all variants under a strict maximum Sharpe performance objective, followed by ENP (L2) at $\alpha=10^1$, both with a modest reduction in net Delta. However, DP (L1) at $\alpha=10^2$ produces a more compelling reduction in net Delta exposure at a moderate coefficient, while simultaneously improving the Sharpe ratio. Stronger penalties achieve substantially narrower net Delta distributions, but at an increasing cost to performance.

Fifth, focusing our analysis on gross Delta, large reductions in gross Delta exposures are attainable, but they come at a significant cost to the risk-adjusted performance of the strategy. DP (L1) at $\alpha=10^4$ delivers the largest reduction in gross Delta, lowering the mean gross position-normalized Delta by approximately $61\%$, but its Sharpe ratio falls below 1.0. In contrast, the highest Sharpe variants reduce mean gross Delta by only approximately $5\%$--$10\%$. Incorporating an intermediate level of regularization such as DP (L1) at $\alpha=10^3$ produces a larger gross reduction of approximately $41\%$, but its Sharpe ratio of $1.753$ is below that of the baseline model.

Overall, the empirical results support a more nuanced picture consistent with the central proposition of this paper. Risk-sensitivity penalties can embed hedging behavior within an end-to-end options strategy without necessarily sacrificing out-of-sample performance. In this case, the scope of our work has focused on the management of Delta, the first-order directional exposure of the options portfolio. At moderate levels, the principal benefit of introducing the risk-sensitivity penalty is a reduction in the persistent net Delta tilt of the portfolio, followed by modest improvements in gross Delta. More aggressive regularization can substantially reduce the Delta exposure of the portfolio, but this comes at an increasingly negative impact on performance. The deterioration in performance observed at extreme values of $\alpha$ confirms that hedging is not costless: once the risk-sensitivity penalty is given overwhelming weight, the model's capacity to generate meaningful trading signals is suppressed as the model deviates from its performance-driven objective. The optimal point therefore depends on whether the portfolio manager prioritizes reducing the net (or gross) Delta exposure of the portfolio or maximizing the risk-adjusted performance of the strategy.

\subsection{Transaction Costs and Turnover Regularization}
\label{subsec:transaction_cost_impact}

In this section, we evaluate the impact of transaction costs on the risk-adjusted performance of each strategy. We measure the daily turnover $\tau_{i, t}$ as the absolute daily change in the scaled trading signal of an individual position:

\begin{equation}
\tau_{i, t} = \sigma_{\text{tgt}} \left| \frac{X_{i, t}}{\sigma_{i, t}} -  \frac{X_{i, t-1}}{\sigma_{i, t-1}} \right|
\end{equation}

\begin{figure*}[htbp]
    \centering
    \includegraphics[width=\textwidth]{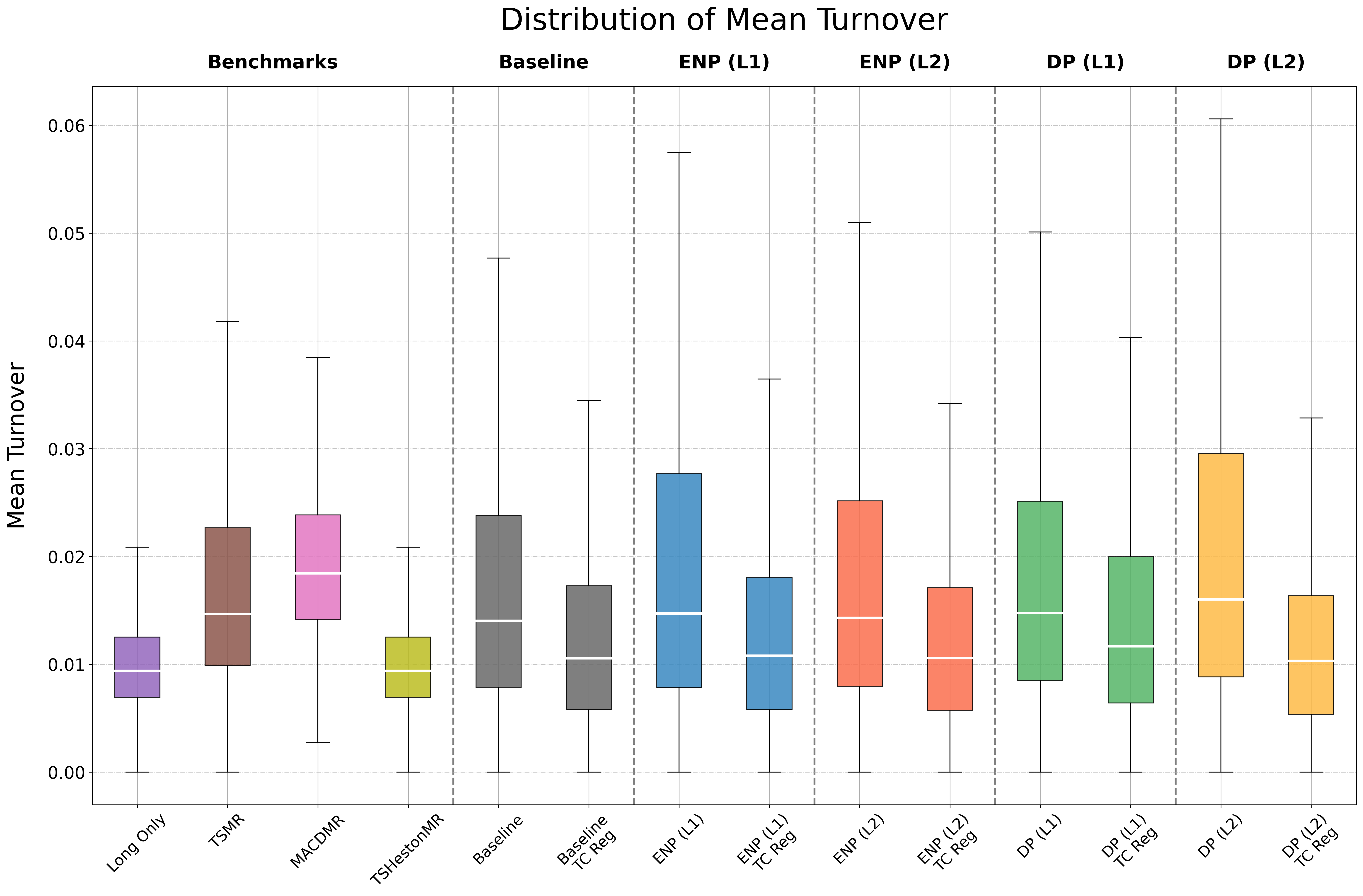}
    \caption{Distribution of Mean Turnover for Strategies}
    \label{fig:mean_turnover_distribution}
\end{figure*}

This turnover metric captures rebalancing activity arising from both changes in the model's signal and fluctuations in the volatility estimates. Figure \ref{fig:mean_turnover_distribution} shows the distributions of the cross-sectional average daily turnover for each strategy.

In practice, rebalancing a systematic portfolio incurs microstructure costs that erode the strategy's returns in proportion to turnover. To quantify this, we define the turnover-adjusted portfolio return as:

\begin{equation}
    \tilde{R}_{t+1}^{\Pi} = \frac{1}{N_t} \sum_{i=1}^{N_t}
    \left(
        X_{i,t} \frac{\sigma^*}{\sigma_{i,t}} \, R_{i, t+1}
        \;-\; c \, \cdot \tau_{i,t}
    \right)
    \label{eq:net_return}
\end{equation}

where $c$ denotes the proportional cost per unit of turnover, expressed in basis points. We then compute the annualized Sharpe ratio of each strategy from the net return series for a range of cost assumptions from $c = 0$ to $100$ bps. Table~\ref{table:sharpe_ratios_tc} reports the results, with all strategies rescaled to target volatility.

We adopt the approach of \cite{lim2019enhancing, tan2023spatio, tan2024deep} and replace the portfolio returns in the training objective with the turnover-adjusted returns defined in Equation~(\ref{eq:net_return}). Concretely, the model is trained to maximize the performance-driven objective of the net return series at a prescribed cost level~$c$, embedding the cost of rebalancing directly into the loss and penalizing signals that generate excessive turnover. Models trained under this modified objective are denoted TC Reg in Table~\ref{table:sharpe_ratios_tc}.

We observe that all baseline and regularized models outperform the benchmark models across the entire range of transaction costs. At the zero-cost level, the performance mirrors the results of Table~\ref{table:best_models}, with DP (L1) achieving the highest Sharpe ratio of 2.339. As transaction costs increase, DP (L1) remains the best-performing model up to 50 bps, achieving a Sharpe ratio of 1.574. At the highest cost level of 100 bps, Baseline LSTM + TC Reg achieves the highest Sharpe ratio of 0.871, followed closely by DP (L1) at 0.860, both outperforming all benchmarks at zero cost. The L1-based penalties consistently outperform their L2 counterparts across both the ENP and DP variants, particularly at higher cost levels. At 100 bps, ENP (L1) achieves a Sharpe ratio of 0.718 compared to 0.634 for ENP (L2), and DP (L1) achieves 0.860 compared to 0.763 for DP (L2). Taken together, these results demonstrate that the combination of risk-sensitivity penalties and turnover regularization provides a robust framework for deploying end-to-end option strategies under realistic trading conditions.

\begin{table*}[htbp]

\centering

\caption{Impact of Transaction Costs on Sharpe Ratio -- Rescaled to Target Volatility}

\label{table:sharpe_ratios_tc}

\resizebox{\textwidth}{!}{

\begin{tabular}{lccccccccccc}

\hline \toprule

\textbf{Model} & \textbf{0.0 bps} & \textbf{0.5 bps} & \textbf{1.0 bps} & \textbf{2.0 bps} & \textbf{3.0 bps} & \textbf{4.0 bps} & \textbf{5.0 bps} & \textbf{10.0 bps} & \textbf{20.0 bps} & \textbf{50.0 bps} & \textbf{100.0 bps} \\

\midrule

\multicolumn{12}{l}{\underline{\textbf{Benchmarks}}} \\

Long Only & 0.621 & 0.620 & 0.618 & 0.616 & 0.613 & 0.611 & 0.608 & 0.596 & 0.571 & 0.496 & 0.372 \\

TSMR & 0.822 & 0.820 & 0.817 & 0.812 & 0.807 & 0.802 & 0.797 & 0.772 & 0.723 & 0.574 & 0.326 \\

MACDMR & 0.785 & 0.781 & 0.776 & 0.767 & 0.758 & 0.750 & 0.741 & 0.697 & 0.608 & 0.343 & -0.100 \\

TSHestonMR & 0.747 & 0.745 & 0.742 & 0.737 & 0.732 & 0.727 & 0.722 & 0.697 & 0.647 & 0.497 & 0.247 \\

CSHestonMR & 0.742 & 0.741 & 0.739 & 0.736 & 0.733 & 0.730 & 0.727 & 0.711 & 0.680 & 0.587 & 0.431 \\

\midrule

\multicolumn{12}{l}{\underline{\textbf{Baseline}}} \\

Baseline LSTM & 2.131 & 2.124 & 2.117 & 2.104 & 2.090 & 2.076 & 2.062 & 1.994 & 1.859 & 1.458 & 0.816 \\

Baseline LSTM + TC Reg. & 1.840 & 1.835 & 1.830 & 1.820 & 1.810 & 1.800 & 1.791 & 1.742 & 1.644 & 1.352 & \textbf{0.871} \\

\midrule

\multicolumn{12}{l}{\underline{\textbf{Regularized Models}}} \\

\multicolumn{12}{l}{\textbf{Exposure-Normalized Penalty}} \\

ENP (L1) & 2.255 & 2.246 & 2.238 & 2.222 & 2.206 & 2.190 & 2.174 & 2.095 & 1.936 & 1.467 & 0.718 \\

ENP (L1) + TC Reg. & 1.932 & 1.926 & 1.921 & 1.909 & 1.897 & 1.885 & 1.874 & 1.815 & 1.698 & 1.348 & 0.770 \\

ENP (L2) & 2.260 & 2.251 & 2.242 & 2.224 & 2.207 & 2.189 & 2.171 & 2.083 & 1.909 & 1.405 & 0.634 \\

ENP (L2) + TC Reg. & 1.788 & 1.783 & 1.778 & 1.769 & 1.759 & 1.750 & 1.740 & 1.692 & 1.596 & 1.310 & 0.839 \\

\midrule

\multicolumn{12}{l}{\textbf{Drift Penalty}} \\

DP (L1) & \textbf{2.339} & \textbf{2.331} & \textbf{2.324} & \textbf{2.308} & \textbf{2.292} & \textbf{2.276} & \textbf{2.261} & \textbf{2.182} & \textbf{2.027} & \textbf{1.574} & 0.860 \\

DP (L1) + TC Reg. & 1.803 & 1.798 & 1.793 & 1.782 & 1.772 & 1.761 & 1.751 & 1.699 & 1.595 & 1.286 & 0.777 \\

DP (L2) & 2.255 & 2.247 & 2.240 & 2.224 & 2.209 & 2.193 & 2.177 & 2.100 & 1.946 & 1.491 & 0.763 \\

DP (L2) + TC Reg. & 1.594 & 1.589 & 1.585 & 1.576 & 1.567 & 1.558 & 1.549 & 1.504 & 1.413 & 1.142 & 0.690 \\

\bottomrule \hline

\end{tabular}

}

\end{table*}

\section{Conclusion} 
\label{conclusion} 

In this paper, we address the challenge of managing complex risk exposures 
in end-to-end deep learning strategies for options trading. We establish that explicitly penalizing portfolio-level risk sensitivities natively embeds hedging behavior into the network's learning process. We introduce a loss function that combines a performance-driven objective with a differentiable risk-sensitivity penalty, and develop two penalty variants -- an exposure-normalized penalty and a Greek-ratio drift penalty. The exposure-normalized penalty measures directional exposure per unit of gross allocation, while the drift penalty identifies contracts whose Greek profile indicates substantial departure from the at-the-money region. Both variants are designed to avoid the degenerate solution admitted by a naive risk-sensitivity penalty, where the optimizer trivially reduces the risk term by uniformly scaling all trading signals towards zero.

We evaluate the framework on static delta-neutral straddle portfolios constructed from Nasdaq 100 equity options spanning over a decade. While an unregularized end-to-end model achieves strong risk-adjusted returns compared to traditional benchmarks, we show that the strategy exhibits a persistent net positive Delta bias. We demonstrate that appropriately calibrated risk-sensitivity penalties can simultaneously improve risk-adjusted returns and reduce directional exposure. The relationship between regularization strength and performance suggests that moderate risk-sensitivity regularization acts as a beneficial inductive bias, steering the model towards learning strategies that achieve a higher risk-adjusted return while simultaneously reducing the portfolio's net directional bias.

A key contribution of this work is the joint analysis of net and gross position-normalized Delta, which provides a crucial distinction between aggregate portfolio tilt and position-level exposure. We find that at moderate regularization strengths, the proposed penalties primarily reduce the portfolio's net directional tilt, with more modest reductions in gross Delta. Among all variants, the L2 variants offer the most favorable trade-off, exhibiting robustness to the choice of risk-aversion coefficient $\alpha$ and maintaining competitive out-of-sample performance across a wide range of regularization strengths.

We further demonstrate that the framework retains its advantages under realistic trading conditions. When evaluated across transaction cost assumptions ranging up to 100 basis points, the regularized models substantially outperformed all benchmarks, with the L2 variants proving to be slightly more resilient to the impact of transaction costs than the L1 variants. Incorporating turnover regularization during training further improved robustness at high cost levels, with the best variant retaining a Sharpe ratio of 1.053 at 100 bps.

Several directions for future work emerge naturally from our framework. First, the risk-sensitivity penalty can be readily extended to target alternative Greeks or risk-sensitivity measures, either individually or jointly, such as incorporating Gamma and Vega, thereby enabling multi-dimensional risk control. Second, since the framework is by construction agnostic to the option pricing model used to evaluate the Greeks, substituting risk measures derived from alternative models requires no modification to the objective function. Third, the application of this methodology to other option structures beyond straddle portfolios, and to broader asset classes, is a promising direction for future research. More generally, the framework naturally extends to other portfolio optimization settings in which undesirable exposures must be minimized, for example, controlling net factor exposures in an equity portfolio. Finally, investigating adaptive or learned schedules for the risk-aversion coefficient $\alpha$, rather than treating it as a fixed hyperparameter, may yield further improvements in balancing the dual objectives of performance and risk control.

\section{Acknowledgements}
We would like to thank the Oxford-Man Institute of Quantitative Finance for providing compute resources.

\bibliographystyle{plain} 
\bibliography{references}

@article{moskowitz2012time,
  title={Time {Series Momentum}},
  author={Moskowitz, Tobias J and Ooi, Yao Hua and Pedersen, Lasse Heje},
  journal={Journal of Financial Economics},
  volume={104},
  number={2},
  pages={228--250},
  year={2012},
  publisher={Elsevier},
doi={https://doi.org/10.1016/j.jfineco.2011.11.003}
}

@article{baz2015dissecting,
  title={Dissecting {Investment Strategies} in the {Cross Section} and {Time Series}},
  author={Baz, Jamil and Granger, Nicolas and Harvey, Campbell R and Le Roux, Nicolas and Rattray, Sandy},
  journal={SSRN 2695101},
  year={2015}
}

@article{tan2023spatio,
  title={{Spatio-Temporal Momentum}: {Jointly Learning Time-Series} and {Cross-Sectional Strategies}},
  author={Tan, Wee Ling and Roberts, Stephen and Zohren, Stefan},
  journal={The Journal of Financial Data Science},
  volume={5},
  number={3},
  pages={107--129},
  year={2023},
  publisher={Portfolio Management Research},
doi={10.3905/jfds.2023.1.130}
}

@article{lim2019enhancing,
  title={Enhancing {Time-Series Momentum Strategies Using Deep Neural Networks}},
  author={Lim, Bryan and Zohren, Stefan and Roberts, Stephen},
  journal={The Journal of Financial Data Science},
  volume={1},
  number={4},
  pages={19--38},
  year={2019},
  publisher={Institutional Investor Journals Umbrella},
doi={10.3905/jfds.2019.1.015}
}

@article{wood2021trading,
  title={Trading with the {Momentum Transformer}: {An Intelligent} and {Interpretable Architecture}},
  author={Wood, Kieran and Giegerich, Sven and Roberts, Stephen and Zohren, Stefan},
  journal={arXiv:2112.08534, Risk},
  year={2023}
}

@article{jegadeesh1993returns,
  title={Returns to {Buying Winners} and {Selling Losers}: {Implications} for {Stock Market Efficiency}},
  author={Jegadeesh, Narasimhan and Titman, Sheridan},
  journal={The Journal of Finance},
  volume={48},
  number={1},
  pages={65--91},
  year={1993},
  publisher={Wiley Online Library},
doi={https://doi.org/10.1111/j.1540-6261.1993.tb04702.x}
}

@article{poterba1988mean,
  title={Mean {Reversion} in {Stock Prices}: {Evidence} and {Implications}},
  author={Poterba, James M and Summers, Lawrence H},
  journal={Journal of Financial Economics},
  volume={22},
  number={1},
  pages={27--59},
  year={1988},
  publisher={Elsevier},
doi={https://doi.org/10.1016/0304-405X(88)90021-9}
}

@article{de1985does,
  title={Does the {Stock Market Overreact}?},
  author={De Bondt, Werner FM and Thaler, Richard},
  journal={The Journal of Finance},
  volume={40},
  number={3},
  pages={793--805},
  year={1985},
  publisher={Wiley Online Library},
doi={https://doi.org/10.2307/2327804}
}

@inproceedings{tan2024deep,
  title={Deep Learning for Options Trading: An End-To-End Approach},
  author={Tan, Wee Ling and Roberts, Stephen and Zohren, Stefan},
  booktitle={Proceedings of the 5th ACM International Conference on AI in Finance},
  pages={487--495},
  year={2024}
}

@article{black1973pricing,
title={The {Pricing} of {Options} and {Corporate Liabilities}},
author={Black, Fischer and Scholes, Myron},
journal={Journal of Political Economy},
volume={81},
number={3},
pages={637--654},
year={1973},
publisher={The University of Chicago Press},
ISSN = {00223808, 1537534X},
URL = {http://www.jstor.org/stable/1831029}
}

@article{merton1973theory,
title={Theory of {Rational Option Pricing}},
author={Merton, Robert C},
journal={The Bell Journal of Economics and Management Science},
number = {1},
pages = {141--183},
publisher = {[Wiley, RAND Corporation]},
volume = {4},
year = {1973},
ISSN = {00058556},
doi={https://doi.org/10.2307/3003143}
}

@article{cox1979option,
title={Option {Pricing}: {A Simplified Approach}},
author={Cox, John C and Ross, Stephen A and Rubinstein, Mark},
journal={Journal of Financial Economics},
volume={7},
number={3},
pages={229--263},
year={1979},
publisher={Elsevier},
issn = {0304-405X},
doi = {https://doi.org/10.1016/0304-405X(79)90015-1},
url = {https://www.sciencedirect.com/science/article/pii/0304405X79900151}
}

@article{buchner2022factor,
  title={A {Factor Model} for {Option Returns}},
  author={B{\"u}chner, Matthias and Kelly, Bryan},
  journal={Journal of Financial Economics},
  volume={143},
  number={3},
  pages={1140--1161},
  year={2022},
  publisher={Elsevier},
doi={https://doi.org/10.1016/j.jfineco.2021.12.007}
}

@article{leland1985option,
  title={Option Pricing and Replication with Transactions Costs},
  author={Leland, Hayne E},
  journal={The Journal of Finance},
  volume={40},
  number={5},
  pages={1283--1301},
  year={1985},
  publisher={Wiley Online Library}
}

@article{boyle1992option,
  title={Option Replication in Discrete Time with Transaction Costs},
  author={Boyle, Phelim P and Vorst, Ton},
  journal={The Journal of Finance},
  volume={47},
  number={1},
  pages={271--293},
  year={1992},
  publisher={Wiley Online Library}
}

@article{figlewski1989options,
  title={Options Arbitrage in Imperfect Markets},
  author={Figlewski, Stephen},
  journal={The Journal of Finance},
  volume={44},
  number={5},
  pages={1289--1311},
  year={1989},
  publisher={Wiley Online Library}
}

@article{buehler2019deep,
author = {Buehler, Hans and Gonon, Lukas and Teichmann, Josef and Wood, Ben},
title = {Deep {Hedging}},
journal = {Quantitative Finance},
volume = {19},
number = {8},
pages = {1271--1291},
year = {2019},
publisher = {Routledge},
doi = {10.1080/14697688.2019.1571683},
}

@article{kolm2019dynamic,
title={Dynamic {Replication} and {Hedging}: {A Reinforcement Learning Approach}},
author={Kolm, Petter N and Ritter, Gordon},
journal={The Journal of Financial Data Science},
volume={1},
number={1},
pages={159--171},
year={2019},
publisher={Institutional Investor Journals Umbrella}, 
doi={10.3905/jfds.2019.1.1.159}
}

@inproceedings{gao2023deeper,
  title={Deeper Hedging: A New Agent-based Model for Effective Deep Hedging},
  author={Gao, Kang and Weston, Stephen and Vytelingum, Perukrishnen and Stillman, Namid and Luk, Wayne and Guo, Ce},
  booktitle={Proceedings of the Fourth ACM International Conference on AI in Finance},
  pages={270--278},
  year={2023}
}

@inproceedings{hirano2023adversarial,
  title={Adversarial deep hedging: Learning to hedge without price process modeling},
  author={Hirano, Masanori and Minami, Kentaro and Imajo, Kentaro},
  booktitle={Proceedings of the Fourth ACM International Conference on AI in Finance},
  pages={19--26},
  year={2023}
}

@inproceedings{mueller2024fast,
  title={Fast deep hedging with second-order optimization},
  author={Mueller, Konrad and Akkari, Amira and Gonon, Lukas and Wood, Ben},
  booktitle={Proceedings of the 5th ACM International Conference on AI in Finance},
  pages={319--327},
  year={2024}
}

@article{hutchinson1994nonparametric,
title={A {Nonparametric Approach} to {Pricing} and {Hedging Derivative Securities Via Learning Networks}},
author={Hutchinson, James M and Lo, Andrew W and Poggio, Tomaso},
journal={The Journal of Finance},
volume={49},
number={3},
pages={851--889},
year={1994},
publisher={Wiley Online Library},
ISSN = {00221082, 15406261},
doi={https://doi.org/10.2307/2329209}
}

@article{ivașcu2021option,
title={Option {Pricing} using {Machine Learning}},
author={Ivașcu, Codruț-Florin},
journal={Expert Systems with Applications},
volume={163},
pages={113799},
year={2021},
publisher={Elsevier},
issn = {0957-4174},
doi = {https://doi.org/10.1016/j.eswa.2020.113799},
url = {https://www.sciencedirect.com/science/article/pii/S0957417420306187}
}

@article{bali2023option,
title={Option {Return Predictability} with {Machine Learning} and {Big Data}},
author={Bali, Turan G and Beckmeyer, Heiner and Moerke, Mathis and Weigert, Florian},
journal={The Review of Financial Studies},
volume={36},
number={9},
pages={3548--3602},
year={2023},
publisher={Oxford University Press}, 
issn = {0893-9454},
doi = {10.1093/rfs/hhad017},
url = {https://doi.org/10.1093/rfs/hhad017},
}

@article{coval2001expected,
title={Expected {Option Returns}},
author={Coval, Joshua D and Shumway, Tyler},
journal={The Journal of Finance},
volume={56},
number={3},
pages={983--1009},
year={2001},
publisher={Wiley Online Library}, 
doi = {https://doi.org/10.1111/0022-1082.00352},
url = {https://onlinelibrary.wiley.com/doi/abs/10.1111/0022-1082.00352}
}

@article{goyal2009cross,
title={Cross-section of {Option Returns} and {Volatility}},
author={Goyal, Amit and Saretto, Alessio},
journal={Journal of Financial Economics},
volume={94},
number={2},
pages={310--326},
year={2009},
publisher={Elsevier},
issn = {0304-405X},
doi = {https://doi.org/10.1016/j.jfineco.2009.01.001},
url = {https://www.sciencedirect.com/science/article/pii/S0304405X09001251}
}

@article{vasquez2017equity,
  title={Equity {Volatility Term Structures} and the {Cross Section} of {Option Returns}},
  author={Vasquez, Aurelio},
  journal={Journal of Financial and Quantitative Analysis},
  volume={52},
  number={6},
  pages={2727--2754},
  year={2017},
  publisher={Cambridge University Press},
doi={https://doi.org/10.1017/S002210901700076X}
}

@article{heston2023option,
  title={Option {Momentum}},
  author={Heston, Steven L and Jones, Christopher S and Khorram, Mehdi and Li, Shuaiqi and Mo, Haitao},
  journal={The Journal of Finance},
  volume={78},
  number={6},
  pages={3141--3192},
  year={2023},
  publisher={Wiley Online Library},
doi={https://doi.org/10.1111/jofi.13279}
}

@book{hull2016options,
  title={Options, futures, and other derivatives},
  author={Hull, John C and Basu, Sankarshan},
  year={2016},
  publisher={Pearson Education India}
}

@article{estrella1997approximation,
  title={Approximation of Changes in Option Values and Hedge Ratios: How Large Are the Errors?},
  author={Estrella, Arturo and Kambhu, John},
  journal={Research Paper, Federal Reserve Bank of New York},
  year={1997}
}

@article{sharpe1998sharpe,
  title={The {Sharpe Ratio}},
  author={Sharpe, William F},
  journal={The Journal of Portfolio Management},
  volume={21},
  number = {1},
  pages={49--58},
  year={1994},
doi={10.3905/jpm.1994.409501}
}

@article{harvey2018impact,
  title={The {Impact} of {Volatility Targeting}},
  author={Harvey, Campbell R and Hoyle, Edward and Korgaonkar, Russell and Rattray, Sandy and Sargaison, Matthew and Van Hemert, Otto},
  journal={The Journal of Portfolio Management},
  volume={45},
  number={1},
  pages={14--33},
  year={2018},
  publisher={Institutional Investor Journals Umbrella},
doi={10.3905/jpm.2018.45.1.014}
}

@article{kingma2014adam,
  title={Adam: {A Method} for {Stochastic Optimization}},
  author={Kingma, Diederik P and Ba, Jimmy},
  journal={arXiv:1412.6980},
  year={2014}
}

@article{zhang2020deep,
  title={Deep {Learning} for {Portfolio Optimization}},
  author={Zhang, Zihao and Zohren, Stefan and Roberts, Stephen},
  journal={The Journal of Financial Data Science},
  volume={2},
  number={4},
  pages={8--20},
  year={2020},
  publisher={Institutional Investor Journals Umbrella}
}

\clearpage
\appendix

\section{Hyperparameter Optimization}
\label{appendix_a}

\begin{table}[H]
\centering
\caption{Hyperparameter Search Range}
\label{table:hyperparameter_search}
\begin{tabular}{lll}
\toprule
\textbf{Hyperparameters}           & \textbf{Search Grid}              \\ \midrule
Minibatch Size                     & 32, 64, 128, 256             \\
Dropout Rate                       & 0.1, 0.2, 0.3, 0.4, 0.5           \\
Hidden Layer Size                  & 5, 10, 20, 40, 80, 160       \\
Learning Rate                      & $10^{-5},~ 10^{-4},~ 10^{-3},~ 10^{-2},~ 10^{-1},~ 10^{0}$                       \\
Max Gradient Norm                  & $10^{-4},~ 10^{-3},~ 10^{-2},~ 10^{-1},~ 10^{0},~ 10^{1}$                       \\
Risk-aversion Coefficient               & $10^{0},~ 10^{1},~ 10^{2},~ 10^{3},~ 10^{4},~ 10^{5},~ 10^{6},~ 10^{7}$                       \\
\bottomrule
\end{tabular}
\end{table}

\section{Time-Series Plots of Net Position-Normalized Delta}
\label{appendix_b}

\begin{figure*}[htbp]
    \centering
    \includegraphics[width=\textwidth]{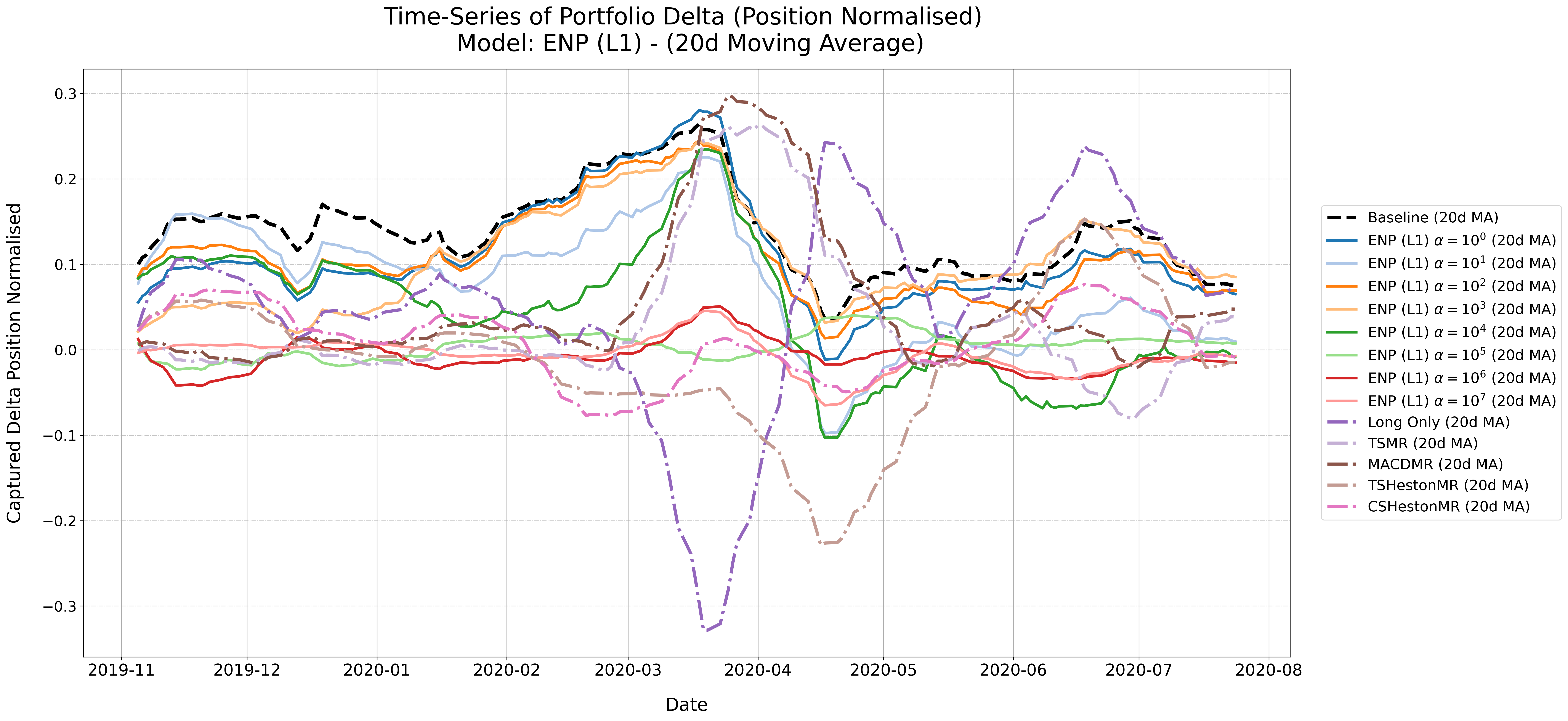}
    \caption{Time-Series Plot of Daily Net Position-Normalized Delta - ENP (L1)}
    \label{fig:timeseries_plot_absposnorm_position_normalised_delta_MA20}
\end{figure*}

\begin{figure*}[htbp]
    \centering
    \includegraphics[width=\textwidth]{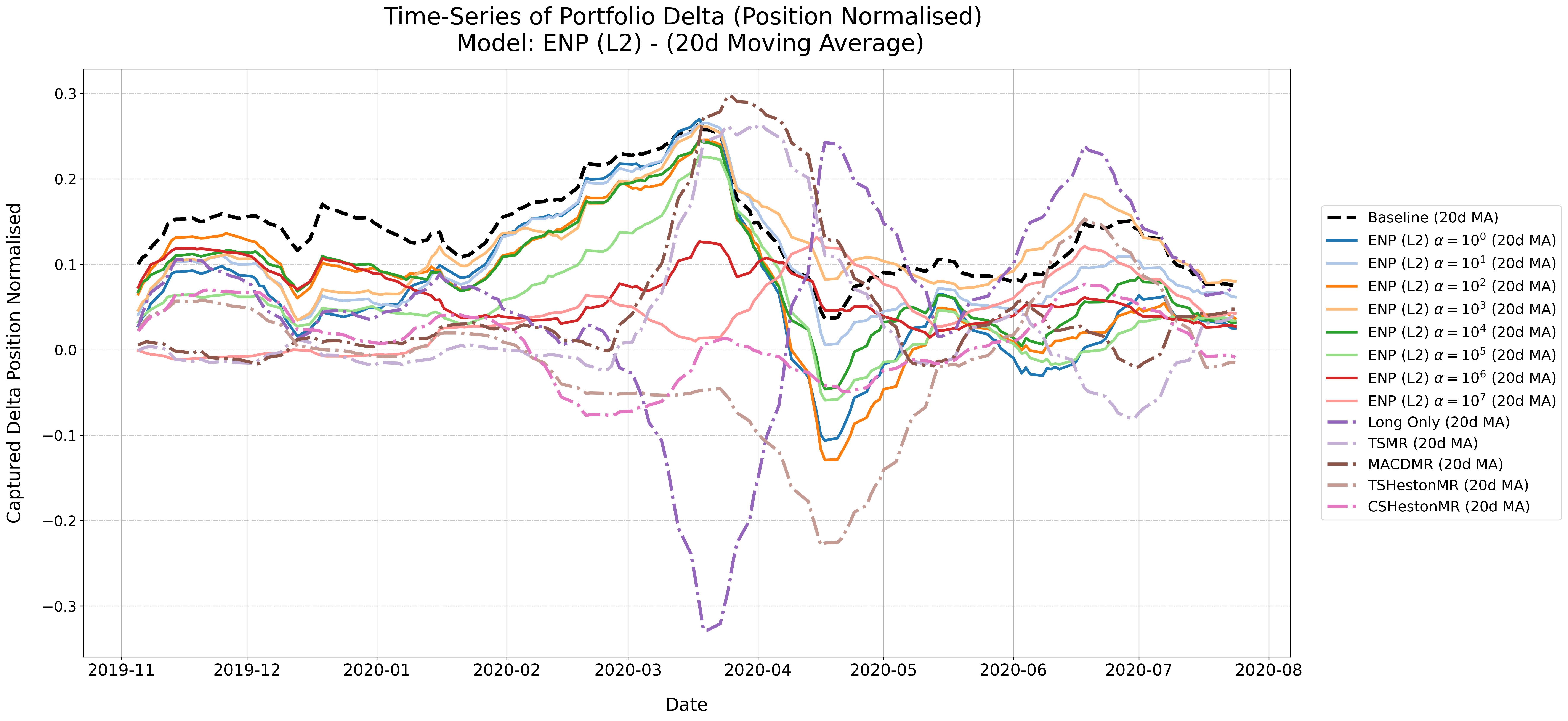}
    \caption{Time-Series Plot of Daily Net Position-Normalized Delta - ENP (L2)}
    \label{fig:timeseries_plot_squposnorm_position_normalised_delta_MA20}
\end{figure*}

\begin{figure*}[htbp]
    \centering
    \includegraphics[width=\textwidth]{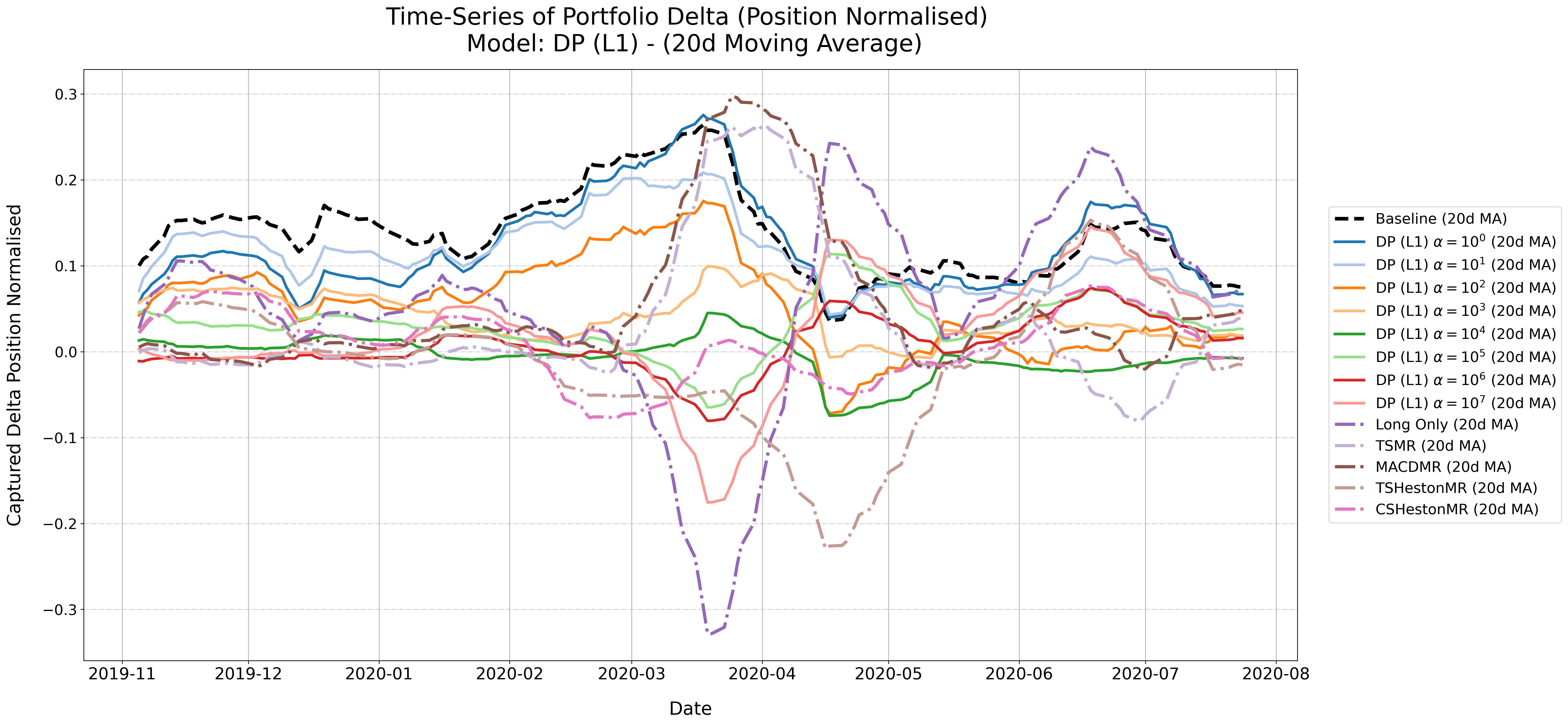}
    \caption{Time-Series Plot of Daily Net Position-Normalized Delta - DP (L1)}
    \label{fig:timeseries_plot_abstrfnorm_6_highreg_position_normalised_delta_MA20}
\end{figure*}

\begin{figure*}[htbp]
    \centering
    \includegraphics[width=\textwidth]{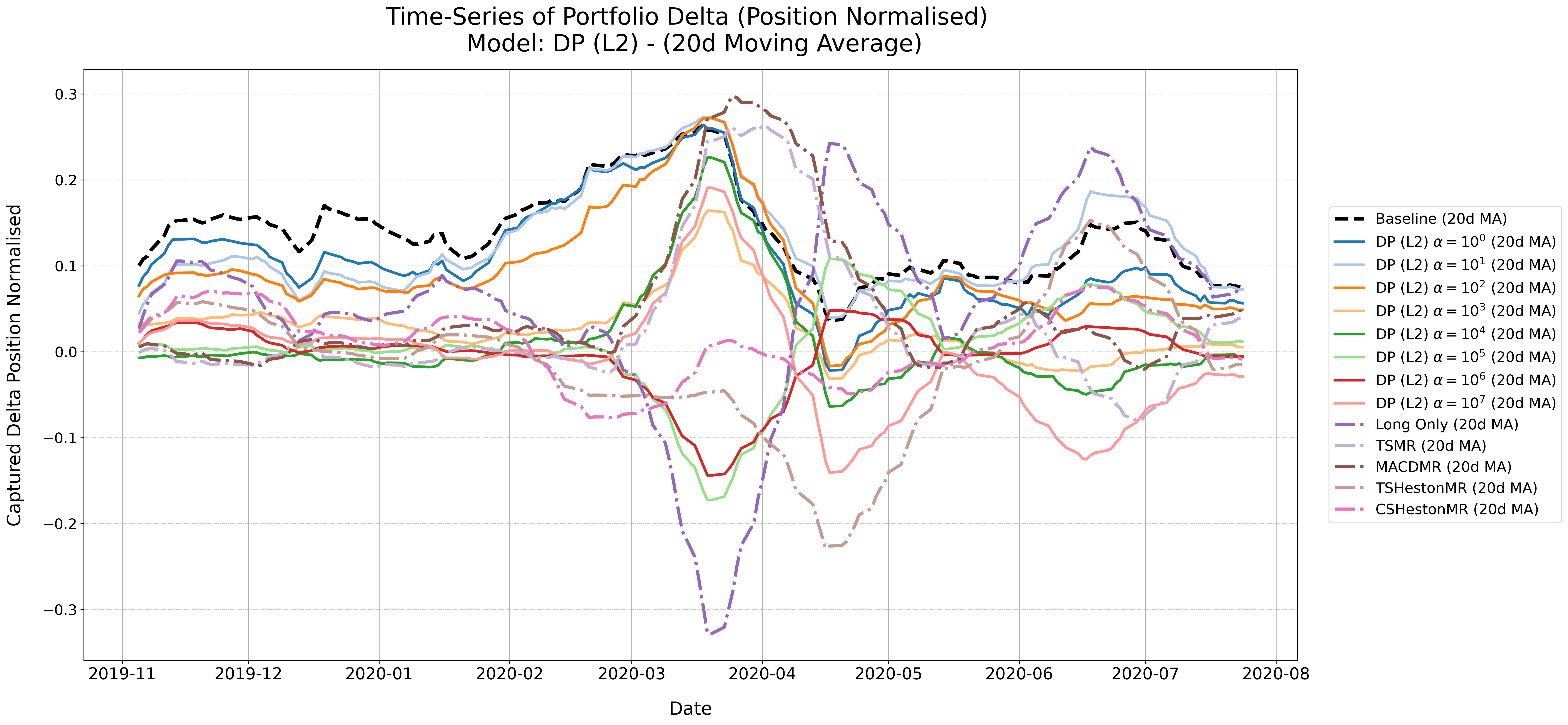}
    \caption{Time-Series of Daily Net Position-Normalized Delta - DP (L2)}
    \label{fig:timeseries_plot_squtrfnorm_6_highreg_position_normalised_delta_MA20}
\end{figure*}

\end{document}